\documentclass[aps,prb,twocolumn,superscriptaddress,floatfix,showpacs]{revtex4-2}
\usepackage[colorlinks,linkcolor={blue},citecolor={blue},urlcolor={blue}]{hyperref} 
\usepackage{graphicx}
\usepackage{bm}
\usepackage[mathlines]{lineno}
\usepackage{graphicx}
\usepackage{amsfonts,amssymb,amsmath}  
\usepackage{float}
\usepackage{siunitx}
\usepackage{multirow}
\usepackage{pifont} % for checkmark 
\usepackage[normalem]{ulem}
\usepackage{wasysym} % for hexagon and kagome symbol
\usepackage[dvipsnames]{xcolor}

\makeatletter
\def\maketitle{
\@author@finish
\title@column\titleblock@produce
\suppressfloats[t]}
\makeatother
\newcommand{\bl}{\boldsymbol}
\newcommand{\be}{\begin{equation}}
\newcommand{\ee}{\end{equation}}
\newcommand{\beqa}{\begin{eqnarray}}
\newcommand{\eeqa}{\end{eqnarray}}

\begin{document}

\title{Propagating edge and interfacial states in corrugated graphene: Robustness and configurability}

\author{Adel Belayadi}
\affiliation{Deparment of Physics, University of Science and Technology Houari Boumediene, Bab-Ezzouar 16111, Algeria}
\affiliation{Ecole Supérieure des Sciences des Aliments et des Industries Agroalimentaires, ESSAIA, El Harrach 16200, Algeria}
\author{Dawei Zhai}
\email{dzhai@hku.hk}
\affiliation{Department of Physics, The University of Hong Kong, Hong Kong, China}
\author{Nancy Sandler}
\affiliation{Department of Physics and Astronomy, Ohio University, Athens, Ohio 45701, USA}

%\date{\today}

\begin{abstract}
Periodically strained graphene provides a versatile platform to realize moiré-like electronic structures. We show that the interplay between strain-induced pseudomagnetic field and a displacement-field-controlled scalar potential enables the formation of isolated narrow bands. Some of the low-energy bands are topological, carrying valley-opposite Chern numbers. Remarkably, despite a vanishing total Chern number, propagating in-gap edge states emerge in various nanoribbon geometries that preserve valley symmetry. We elucidate the distinct mechanisms responsible for edge states in the zero-energy and higher-energy gaps and demonstrate their robustness against disorders, despite lacking conventional topological protection. Leveraging these properties, we propose device architectures in which a displacement field switches the zero-energy gap and its associated edge channels on and off. Furthermore, split-gate geometries generate topological interfacial states that coexist with the edge modes and can be spatially reconfigured. These results establish strain superlattices as a powerful platform for engineering topological electronic states and electronic transport.
\end{abstract}

\maketitle

\emph{\textbf{Introduction}---}
Van der Waals superlattices created by vertical stacking of two-dimensional materials offer a versatile platform for exploring topological and correlated behavior in low-dimensional systems. Canonical examples include investigations of twisted bilayer graphene and transition metal dichalcogenides, where the development of moir\'e potentials has given rise to topological narrow bands and landmark discoveries such as superconductivity and fractional quantum anomalous Hall states~\cite{TBGCaoYuan2018a,FCIMoTe2Jiaqi2023,FCIMoTe2ShanJie2023,FCIMoTe2Park2023,FCIMoTe2PRX2023}. 
However, despite stringent protocols and remarkable progress in various fabrication methods, accurately controlling the twist angle between the constituent materials—which defines the resulting periodic potentials—remains a core challenge for exploring distinct physics from small to large periods~\cite{JeanieLauNature2022,MoireReviewAdvMater2021a,MoireReviewAdvMater2021b,MoireReviewACSNano2021,FabVdWLiuYu2023,MoireReviewChemRev2024a,MoireReviewChemRev2024b,MoireReviewACSApplMater2024}. 

An alternative strategy for controllably fabricating superlattices relies on using engineered substrates with predefined periodic patterns that can transfer specific potentials to the low-dimensional materials placed on top~\cite{YuhangNL2017,Nadya_npj2D2018,NadyaAPL2019,JinhaiNature2020,Jinhai2026,PeetersPeriodicStrainPRB2020,Antonio2DMater2021a,Antonio2DMater2021b,PhongPRL2022,GraphenePeriodicPseudoBJApplPhys2022,EslamPeriodicStrainPRL2023,TareqNanoLett2023,PeriodicStrainSunKaiPRB2023,PeriodicStrainNetworkModelPRB2023,ZhaiSUSY2024,PeriodicStrainRelaxationPRB2024,NadyaNanoLett2025,Vortexibility2023,Vortexibility2025,PeriodicStrainFCIPRB2025,PeriodicStrainScreeningPRL2025,PeriodicStrainYehSciAdv2020,PeriodicStrainNatCommun2021,PeriodicStrainYakobsonNanoLett2022,PeriodicStrainYakobsonNanoCommun2022,SunKaiStrainPRL2023,SubstrateCoryDean2018}.
In these structures, a designed deformation induces a strain field, usually described in terms of an out-of-plane pseudomagnetic field,
enabling useful analogies with the electron dynamics in an external magnetic field. Importantly, these strain superlattices can be designed to realize narrow Chern bands~\cite{EslamPeriodicStrainPRL2023,TareqNanoLett2023,PeriodicStrainNetworkModelPRB2023,PeriodicStrainSunKaiPRB2023,SunKaiStrainPRL2023,ZhaiSUSY2024} that potentially exhibit fractional quantum anomalous Hall effects~\cite{EslamPeriodicStrainPRL2023,SunKaiStrainPRL2023,ZhaiSUSY2024,Vortexibility2023,Vortexibility2025,PeriodicStrainFCIPRB2025}.

Although two-dimensional strain superlattices with periodicity similar to those found in conventional moir\'e materials (i.e., of the order of $\mathcal{O}(10)$~nm) are promising platforms for exploring both single-particle and correlation-driven phenomena~\cite{SubstrateCoryDean2018}, their experimental realization and systematic investigation have so far remained relatively limited~\cite{JinhaiNature2020,Jinhai2026,NadyaNanoLett2025,Nadya_npj2D2018,NadyaAPL2019}. A major challenge is the controlled fabrication of substrates with the required structural and geometric features. Recent experimental advances have made significant progress in this direction, bringing such systems within reach~\cite{PeriodicStrainFabricateNolan2026,PeriodicStrainFabricate2026}. These new developments also create an opportunity to examine whether strain superlattices offer distinct advantages over moir\'e superlattices that have been studied more extensively.

To address this issue, we present here one example---using graphene as a model material---that capitalizes on the corrugated surface morphology to induce a strain superlattice.
An out-of-plane deformation in the presence of an external perpendicular displacement field gives rise to a periodic scalar potential that is commensurate with the superlattice geometry and modifies the electron dynamics induced by the pseudomagnetic field. As we show below, the combination of both fields introduces novel bulk behavior as manifested by the emergence of an effective staggered mass $\Delta_{\rm eff}$ that gaps out the Dirac points~\cite{SnymanPRB2009}. The magnitude and sign of $\Delta_{\rm eff}$ are tunable by the externally applied displacement field, which helps to control the topology of low-energy bands.

Additionally, for finite-size samples, we find that the interplay of the fields profoundly reshapes the edge-state spectral properties, leading to propagating edge modes with energies within the zero-energy gap. These time-reversal symmetric states are localized on the same boundary of the sample and flow in opposite directions. We also provide new insight into the role of the strain-induced staggered pseudomagentic field whose magnetic flux per unit cell vanishes, and whose influence on the edge-state properties of strained graphene remains incompletely understood~\cite{EdgeStateLocationPRB2017}.
Remarkably, we find that these edge states are robust against various types of edge disorders, making them promising candidates for device applications. We further demonstrate that their presence can be electrically switched on/off through the introduction of an external staggered potential.
Moreover, by placing the superlattice in a split-gate geometry that creates interfaces between regions subjected to opposite displacement fields, we reveal the emergence of propagating interfacial states in addition to the edge states. This architecture enables the realization of reconfigurable electronic pathways, in which the location, number and activation of the interfacial channels can be precisely controlled through the various split gates.

\emph{\textbf{Model---}}
We perform tight-binding calculations on nanoribbon geometries to characterize the properties of edge states, the details of which are provided in the Supporting Information (SI). At low energies, an effective Hamiltonian given by a continuum model can shed light on the electronic and topological properties of the bulk. In the ($\mathcal{A},\,\mathcal{B}$) sublattice basis, it reads
\begin{equation}
	H_{\tau}=v\bl{\sigma}_{\tau}\cdot[\bl{p}+e\tau\bl{A}(\bl{r})]+V(\bl{r})+\Delta\sigma_z,
	\label{Eq:Hamiltonian}
\end{equation}
where $v$ is the Fermi velocity of graphene, $\tau=\pm$ is the valley index, $\bl{\sigma}_{\tau}=(\tau\sigma_x,\,\sigma_y)$ is composed of Pauli matrices, $\tau\bl{A}(\bl{r})$ denotes the periodic strain-induced pseudovector potential, $V(\bl{r})$ is a periodic external scalar potential, $-e<0$ is the electric charge, and $\Delta$ is a constant staggered potential. $\bl{A}(\bl{r})$ and $V(\bl{r})$ have a spatial periodicity $L$ ($\gg$ graphene lattice constant), whose spatial details are specified below. Notice that the pseudovector potential $\tau\bl{A}(\bl{r})$ gives rise to a pseudomagnetic field $\tau\bl{B}(\bl{r})=\tau\nabla\times\bl{A}(\bl{r})$.

%---------------------- Fig.1 --------------------
\begin{figure}[t]
	\centering
	\includegraphics[width=3.4in]{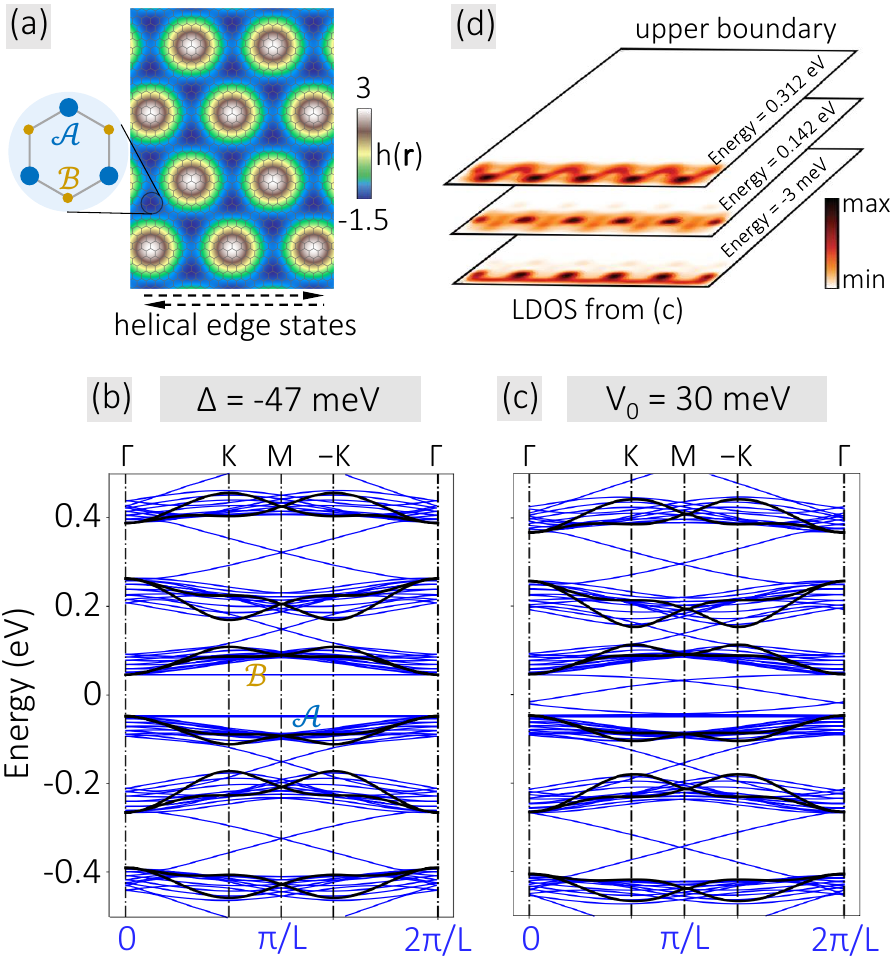}
	\caption{\textbf{Edge states in a periodically strained graphene nanoribbon with zigzag-zigzag boundaries}. (a) Profile of $h(\bl{r})$ that determines both the pseudomagnetic field $B(\bl{r})=B_0h(\bl{r})$ and scalar potential $V(\bl{r})=V_0h(\bl{r})$ profiles. The inset illustrates the $\mathcal{A}$ sublattice polarization with larger blue dots caused by the pseudomagnetic magnetic field in the $\varhexagon$ region (see text). (b) Energy bands of a periodically strained graphene nanoribbon with a staggered potential $\Delta \sigma_z$. The black curves represent bulk energy bands with high symmetry points labeled on the top. The two exact flat bands are $\mathcal{B}$ and $\mathcal{A}$ sublattice polarized respectively. (c) Similar to those in panel (b), but with a periodic scalar potential $V(\bl{r})$. The energy bands are shifted downward by 18~meV for comparison. (d) Local density of states (LDOS) of selected edge states from (c) concentrated near the edge terminated with the $\mathcal{B}$ sublattice (see text). $B_0\approx910$~T in all panels.}~\label{Fig:1}
\end{figure}
%-------------------- end Fig.1--------------------

The low-energy properties can be understood by considering $\mathcal{H}_{\tau}=v\bl{\sigma}_{\tau}\cdot(\bl{p}+e\tau\bl{A})$ as the pristine Hamiltonian and treating $V(\bl{r})$ and $\Delta\sigma_z$ perturbatively~\cite{SnymanPRB2009}. We illustrate this by considering $\mathcal{H}_{+}$ explicitly. $\mathcal{H}_{+}$ has two zero-energy modes at $\bl{k}=0$, i.e., $\psi^{+}_{0}=c_{+}(e^{\phi},\,0)^T$ and $\psi^{-}_{0}=c_{-}(0,\,e^{-\phi})^T$, where $c_{\pm}$ are real normalization constants and $\phi$ is fixed using the Coulomb gauge, i.e., $e\bl{A}/\hbar=-\nabla\times(\phi\hat{z})$ with $\partial_{\bl{r}}^{2}\phi=eB/\hbar$. For $\bl{k}\approx0$, in the presence of $V(\bl{r})$ and $\Delta\sigma_z$, degenerate perturbation theory yields an effective Hamiltonian 
\begin{equation}
	\begin{aligned}
		H_{+}^{\rm eff}&=\hbar\bar{v}\bl{\sigma}_{+}\cdot\bl{k}+(\Delta+\Delta_{\rm eff})\sigma_z\\
		\Delta_{\rm eff}&=\frac{1}{2}\int_{\rm uc} V(\bl{r})\left(c_+^2e^{2\phi}-c_-^2e^{-2\phi}\right)d\bl{r}
	\end{aligned},~\label{Eq:Heff}
\end{equation}
where $\bar{v}$ is a renormalized Fermi velocity, $\Delta_{\rm eff}$ is an effective staggered potential integrated over a unit cell, and a constant shift of the total energy has been discarded.
Here, we identify the crucial role that the spatially varying $V(\bl{r})$ plays in bulk properties: its weighted average acts as an effective staggered potential $\Delta_{\rm eff}\sigma_z$ that opens a gap at the Dirac point.
A pseudomagnetic field, which preserves the time-reversal symmetry, leads to identical $\Delta_{\rm eff}$ in the two valleys. In general, $\Delta_{\rm eff}\ne0$, with a magnitude determined by the profiles of $B(\bl{r})$ and $V(\bl{r})$. Its value becomes largest when $B(\bl{r})$ and $V(\bl{r})$ have identical profiles~\cite{SnymanPRB2009,GrapheneGapTonyLowPRB2011}.
The above discussions show that the low-energy dispersions of $H_{\tau}$ are described by gapped Dirac cones; while at high energies, a strong periodic magnetic field without two-fold rotational $C_{2z}$ symmetry can lead to a series of gapped bands~\cite{JinhaiNature2020,PeetersPeriodicStrainPRB2020,Antonio2DMater2021a,Antonio2DMater2021b,PhongPRL2022,EslamPeriodicStrainPRL2023,TareqNanoLett2023,PeriodicStrainSunKaiPRB2023,ZhaiSUSY2024,PeriodicStrainNetworkModelPRB2023,Jinhai2026}.

\emph{\textbf{A simplified model setup---}}
Spatial variations of the fields are represented by functions $B(\bl{r})=B_0f(\bl{r})$ and $V(\bl{r})=V_0h(\bl{r})$, where $B_0$ and $V_0$ are constants characterizing the intensities, and $f(\bl{r})$ and $h(\bl{r})$ describe periodic spatial dependencies to be specified for different cases.
Here $V(\bl{r})$ represents the electrostatic potential on the corrugated surface induced by an out-of-plane displacement field $\mathcal{D}$ from external gates. Its magnitude is $V_0=e\mathcal{D}z_0$ for a graphene height profile $z(\bl{r})=z_0h(\bl{r})$. The displacement field can reach the order of $\mathcal{O}(0.1)$~V/nm in typical two-dimensional materials~\cite{CaoYuanTwistedDoubleBilayer2020,IQAHinMoTe2WSe22021}. For $z_0\sim1$~nm and $\mathcal{D}\sim0.01$~V/nm, one finds $V_0\sim10$~meV. 
Generally, the pseudomagnetic field and scalar potential are expected to have different profiles with $f(\bl{r})\ne h(\bl{r})$. To illustrate the main findings and following Refs.~\cite{JinhaiNature2020,Antonio2DMater2021a,Antonio2DMater2021b,EslamPeriodicStrainPRL2023}, here we consider a simplified model with $f(\bl{r})=h(\bl{r})$, where $h(\bl{r})=\sum_{i=1}^{3}\cos(\bl{b}_{i}\cdot\bl{r})$ characterizes the height corrugation, $\bl{b}_{i}=C_{3z}^{i-1}\bl{b}_1$ and $\bl{b}_1=\frac{4\pi}{\sqrt{3}L}\hat{y}$. 
Alternative models with more realistic profiles $f(\bl{r})\ne h(\bl{r})$ are discussed in Sec.~S2 and Sec.~S3 of the SI, where we show qualitative and quantitative agreement with the results obtained with the simplified model. These results provide strong support for the robustness of our findings in a wide range of different experimental realizations~\cite{JinhaiNature2020,Jinhai2026}.

The simplified model is characterized by $f(\bl{r})=h(\bl{r})$, whose profile is shown in Fig.~\ref{Fig:1}(a). It contains an interconnected hexagonal part (the blue area, denoted by $\varhexagon$) and isolated circular islands (the gray area, denoted by $\bullet$). The $\varhexagon$ regions are particularly crucial for the following discussions. In this model, $B_0>0$ and $V_0>0$ lead to $\Delta_{\rm eff}<0$ [cf. Eq.~(\ref{Eq:Heff})].

We proceed to characterize the electronic and topological properties of a periodically strained graphene membrane. To reduce computational cost in the tight-binding calculations, a small period $L\approx5$~nm is employed, thus an artificially large value for $B_0$ is used to achieve narrow energy bands for a better visualization of the band gaps. $B_0$ can be significantly reduced in realistic configurations with larger $L$ (see Sec.~S1~B of SI). To identify the effects of $V(\bl{r})$ and $\Delta\sigma_z$ separately, we set $V_0=0$ or $\Delta=0$ at a time. The particular case of both $V_0\ne0$ and $\Delta\ne0$ will  be discussed in Fig.~\ref{Fig:3}. Naively, one might expect $V(\bl{r})$ and $\Delta\sigma_z$ to yield qualitatively identical results  based on the perturbation theory arguments [see Eq.~(\ref{Eq:Heff})]. However, in graphene nanoribbons of finite width, we show that $V(\bl{r})$ has nontrivial effects on the in-gap propagating edge states around the charge neutrality point, which cannot be obtained with $\Delta\sigma_z$ alone.

%=========================================
\emph{\textbf{Bulk properties and in-gap propagating edge states---}}
Fig.~\ref{Fig:1}(a) shows a schematic representation of the pseudomagnetic field profile induced by an underlying periodic deformation on graphene as described above. The resulting bulk band structure with either a staggered potential $\Delta\sigma_z$ or a periodic scalar potential $V(\bl{r})$ is depicted by the black curves in Figs.~\ref{Fig:1}(b, c), obtained from tight-binding calculations for a torus geometry [cf. the continuum model Eq.~(\ref{Eq:Hamiltonian})]. One observes that (i) a gap opens around the zero energy in Fig.~\ref{Fig:1}(c), (ii) the black curves in Fig.~\ref{Fig:1}(b) and Fig.~\ref{Fig:1}(c) are similar, confirming that $V(\bl{r})$ indeed behaves as an effective staggered potential $\Delta_{\rm eff}\sigma_z$ in the bulk [see Eq.~(\ref{Eq:Heff})].
The finite $\Delta<0$ [in Fig.~\ref{Fig:1}(b)] or $\Delta_{\rm eff}<0$ [in Fig.~\ref{Fig:1}(c)] and pseudomagnetic fields in the $\pm K$ valleys lead to nontrivial topology in the first valence bands characterized by Chern numbers $C_{v1}^{K}=-1$ and $C_{v1}^{-K}=1$~\cite{EslamPeriodicStrainPRL2023,TareqNanoLett2023,ZhaiSUSY2024}. Notice that these values coincide with the sign of the $\varhexagon$ part of the $\tau B(\bl{r})$ field [see Fig.~\ref{Fig:1}(a)]~\cite{ZhaiSUSY2024,TareqNanoLett2023}. The other energy bands shown in the energy window of Figs.~\ref{Fig:1}(b, c) are topologically trivial. In addition, the pseudomagnetic field polarizes the sublattice pseudospin~\cite{AlexNanoLett2017}, leading in this case to the $\mathcal{A}$ ($\mathcal{B}$) sublattice polarization in the $\varhexagon$ ($\bullet$) region [Fig.~\ref{Fig:1}(a) inset].

%%---------------------- Table --------------------%%
\begin{table}[t]
	\caption{Summary of nanoribbon edge configurations and whether in-gap propagating edge states exist (\ding{51}) or not (\ding{55}).}
	\label{Table:BoundaryConfiguration&EdgeState}
	\begin{tabular}{c|c|c|c|c}
		\hline
		\textbf{upper edge} & zigzag \ding{55} & bearded \ding{51} & bearded \ding{51} & zigzag \ding{55} \\ \hline
		\textbf{lower edge} & zigzag \ding{51} & bearded \ding{55} & zigzag \ding{51} & bearded \ding{55}\\ 
		\hline
	\end{tabular}
\end{table}
%%---------------------- End Table --------------------%%

%%---------------------- Fig.2 --------------------
%%%%%
\begin{figure}[t]
	\centering
	\includegraphics[width=3.4in]{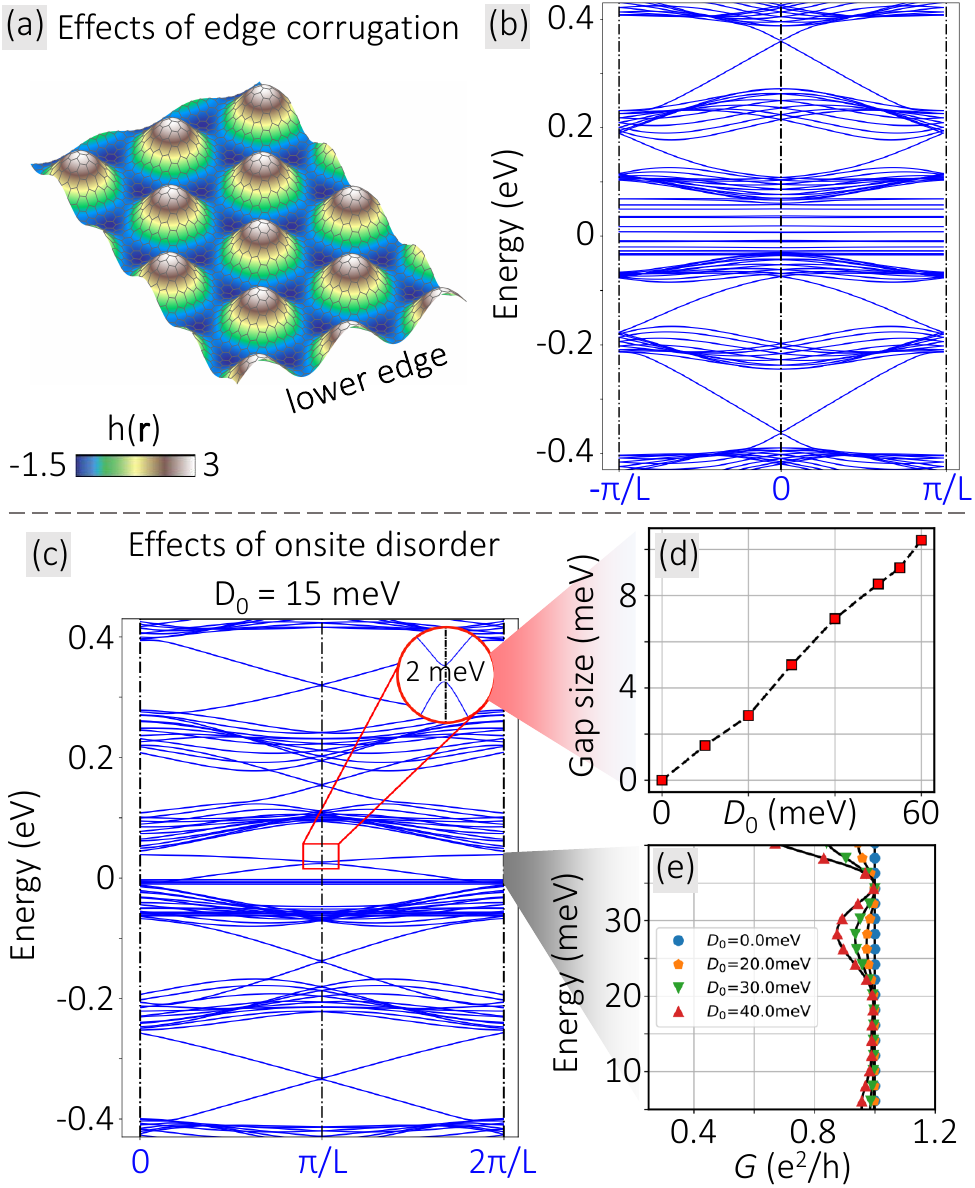}
	\caption{\textbf{Robustness of the edge states}. (a, b) Effects of pseudomagnetic field $B(\bl{r})=B_0h(\bl{r})$ and scalar potential $V(\bl{r})=V_0h(\bl{r})$ profile at the ribbon boundary on the edge state dispersion. An example of an unfavorable boundary termination is shown, where the ribbon's lower edge cuts through the maxima of $h(\bl{r})$. (c--e) Robustness of edge state dispersion and conductance against onsite potential disorder. The favorable edge configuration in Fig.~\ref{Fig:1}(a) is considered. $B_0\approx910$~T, $V_0=30$~meV and $\Delta=0$ in all panels.}~\label{Fig:2}
\end{figure}
%%%%%%%
%%-------------------- end Fig.2--------------------

The blue curves in Figs.~\ref{Fig:1}(b, c) represent the energy bands of a graphene nanoribbon with zigzag termination at both boundaries. 
In addition to affecting the bulk properties by opening a gap at the Dirac point, $V(\bl{r})$ modulates the onsite potential at the boundaries, leading to dispersive edge states in the zero-energy gap (Sec.~S1~C of SI). Importantly, this key feature cannot be achieved with a staggered potential $\Delta\sigma_z$. The ribbon bulk bands (dense blue) are consistent with the bulk (black) bands of a membrane without edges, as both exhibit similar bandwidths. Additionally, a pair of in-gap propagating edge states emerge at various higher energies, which reside at the ribbon's lower boundary [Fig.~\ref{Fig:1}(d)] 
%terminated by $\mathcal{B}$ sublattice atoms [cf. Fig.~\ref{Fig:1}(a)] 
and originate in the two valleys, flowing in opposite directions. It has been shown that these higher-energy edge states can emerge in the presence of only $\tau B(\bl{r})$, i.e., without $\Delta\sigma_z$ and $V(\bl{r})$ (see SI Fig.~S4)~\cite{PhongPRL2022}, similar to edge states in the moir\'e band gaps of twisted materials~\cite{LiuJianpengPRB2019,EdgeStatesTBG2023,ZhangYang1LL2025,EdgeStateAnXingtao,EdgeStateMoTe22026}. 
Intriguingly, we find that the effects of $\Delta\sigma_z$ and $V(\bl{r})$ in a ribbon are  characterized by remarkably distinct behaviors of the zero-energy gap states: while two flat bands appear at the band edges with $\pm\Delta$ in Fig.~\ref{Fig:1}(b), propagating edge states emerge in Fig.~\ref{Fig:1}(c) with $V(\bl{r})$. 
This difference is rooted in the onsite potential modulation at the ribbon's boundaries introduced by the scalar potential $V(\bl{r})$ (see SI Sec.~S1~C).
We also note that the presence/absence of these edge states in the various gaps depend on the edge configuration of the ribbon, as summarized in Table~\ref{Table:BoundaryConfiguration&EdgeState} and Fig.~S4 for the configurations with negligible intervalley scattering. For example, edge states can reside at the ribbon's boundary terminated by sublattice $\mathcal{B}$ atoms [cf. Fig.~\ref{Fig:1}(a)], i.e., the sublattice opposite to the dominant sublattice in the $\varhexagon$ region due to $\tau B$ field-induced sublattice polarization~\cite{EdgeStateLocationPRB2017}. Hence, the pseudomagnetic field in the $\varhexagon$ region not only is crucial for band topology, but it also determines the boundary location of the propagating edge states [also see discussions in Figs.~\ref{Fig:2}(a, b)].

The indistinguishability of bulk states (both the black and the dense blue curves) with identical band topology shown in Figs.~\ref{Fig:1}(b) and (c) contrasts with the absence or presence of edge states in their zero-energy gaps. In addition, the presence of these states depends on the boundary termination, as summarized in Table~\ref{Table:BoundaryConfiguration&EdgeState}. These results are manifestations that the zero-energy propagating edge states are not topological and the corresponding valley Chern numbers cannot be used to determine their number inside the gaps~\cite{ValleyBulkEdgeMorpurgoPRB2010}.
Surprisingly, however, these edge states appear to be quite robust against boundary termination variation and disorder, as we show below, holding potential for device applications. 

In the following, we focus on the edge states shown in Fig.~\ref{Fig:1}(c) that are experimentally relevant at low doping, and consider the zigzag-zigzag (ZZ) boundary configuration (cf. Table~\ref{Table:BoundaryConfiguration&EdgeState}).

%==========================================
\emph{\textbf{Robustness of edge states--}}
Given a graphene ribbon with fixed boundary configuration (i.e., ZZ), the profile of $B(\bl{r})$ and $V(\bl{r})$ at the ribbon boundary (or equivalently, the lattice corrugation at the boundary) affects the robustness of the in-gap propagating edge states. Fig.~\ref{Fig:2}(a) illustrates this by considering a ribbon where the lower boundary (that hosts the edge states) cuts through the maxima intensity regions for $B(\bl{r})$ and $V(\bl{r})$ (centers of gray spots), where the fields' magnitudes have the strongest variations. One observes that the edge state dispersion in the zero-energy gap breaks into a series of flat bands, whereas the edge states in the other higher energy gaps are not affected. This suggests that to ensure the existence of propagating edge states in the zero-energy gap, a quasi-uniform, softly varying field at the boundary is preferred. 
Notice, incidentally, that similar variations in the fields' profile at the upper boundary do not affect the nature of the edge states in the lower boundary, as expected.

Disorder in the nanoribbon can also affect the robustness of edge states. To quantify the effect of disorder, we consider a distribution of onsite random potentials at the zigzag boundaries with intensity in the range of $[-D_0,D_0]$. The zero-energy gap edge states' dispersion remains nearly intact, with the opening of a small gap of size $\ll D_0$ due to intervalley scattering for disorder intensities $D_0$ of the order of $\mathcal{O}(100)$ meV [Figs.~\ref{Fig:2}(c, d)]. The robustness against onsite disorder is also corroborated by the nearly quantized transport conductance ($G\sim 0.9e^2/h$) contributed by these edge states with values of up to $D_0=40$ meV [Fig.~\ref{Fig:2}(e)]. We find that edge states in higher energy gaps are even less affected. In Sec.~S1~E of SI, the effects of lattice vacancies/defects on the ribbon boundaries are also examined.

%%---------------------- Fig.3 --------------------
%%%%%
\begin{figure}[t]
	\centering
	\includegraphics[width=3.4in]{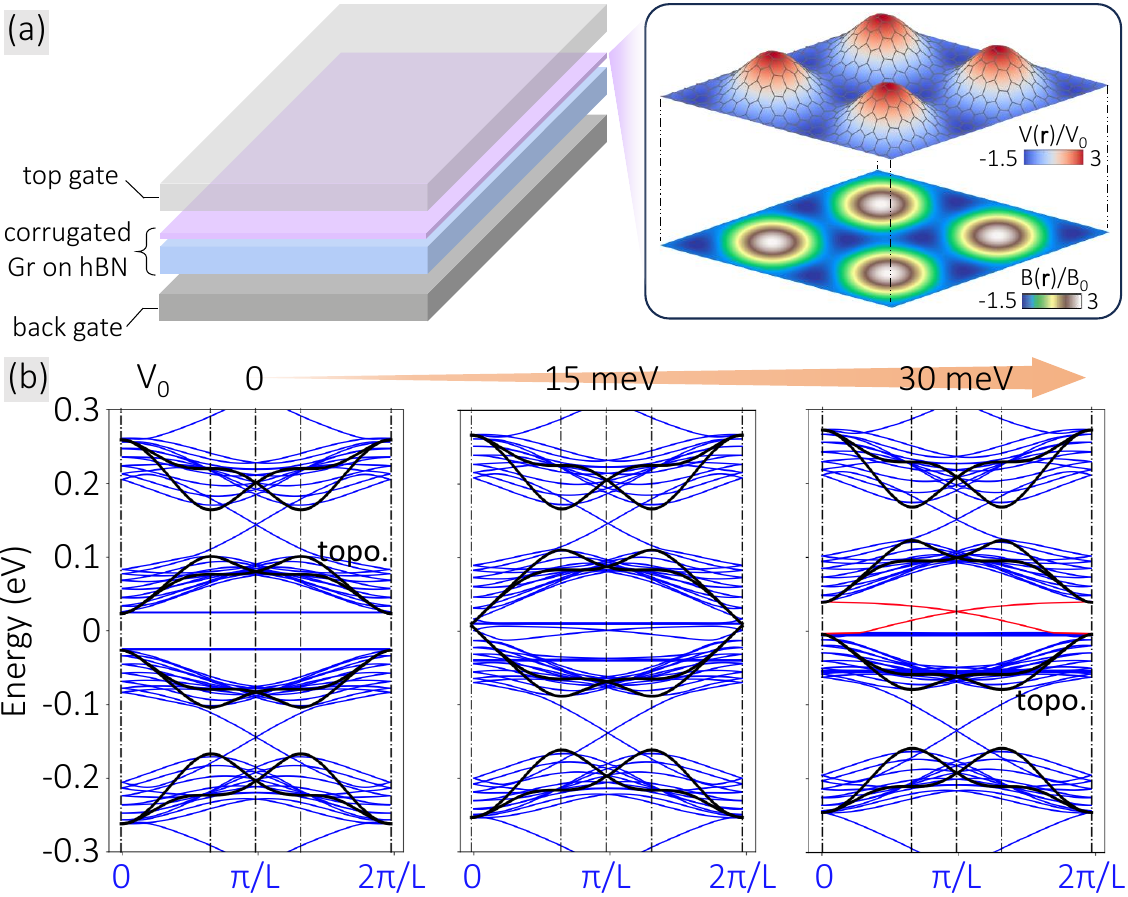}
	\caption{\textbf{Device 1: electrical switch of edge states and band topology}. (a) Schematics of periodically corrugated graphene on hBN with height profile $h(\bl{r})$. An out-of-plane displacement field can be applied to generate a periodic scalar potential $V(\bl{r})=V_0h(\bl{r})$. The hBN introduces a staggered potential $\Delta\sigma_z$ onto graphene, and the pseudomagnetic field reads $B(\bl{r})=B_0h(\bl{r})$. (b) Evolution of the energy bands of a zigzag nanoribbon (blue curves) with respect to $V_0$ by tuning the displacement field. The black curves represent the bulk bands along the same high symmetry path as in Figs.~\ref{Fig:1}(b, c). $B_0\approx910$~T and $\Delta=25$~meV have been used.}~\label{Fig:3}
\end{figure}
%%%%%%%
%%-------------------- end Fig.3--------------------

%%---------------------- Fig.4 --------------------
%%%%%
\begin{figure}[h!]
	\centering
	\includegraphics[width=3.4in]{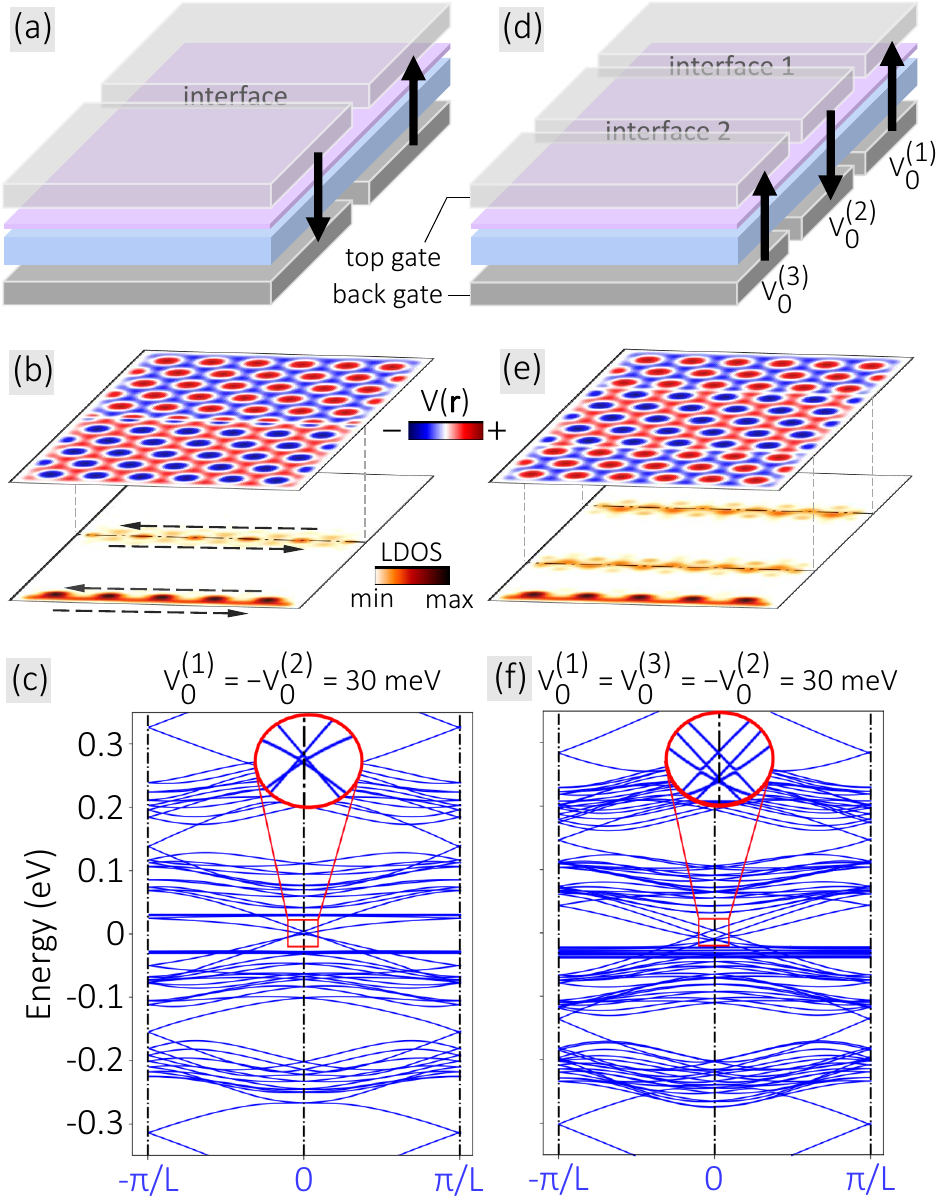}
	\caption{\textbf{Device 2: electrical patterning of edge state arrays}. (a) Schematics of a corrugated zigzag graphene nanoribbon with an interface formed by opposite out-of-plane displacement fields (black arrows). (b) Upper plot shows the distribution of the opposite periodic scalar potentials $V_1(\bl{r})=V_0^{(1)}h(\bl{r})$ and $V_2(\bl{r})=V_0^{(2)}h(\bl{r})$ on the two sides of the interface, where $V_0^{(1)}=-V_0^{(2)}>0$. Lower plot shows the LDOS of the propagating edge states at energy $15$~meV in the zero-energy gap. (c) Energy bands for the interface in (b). The zoom-in shows two pairs of edge states, including the counterflows at the lower boundary of the ribbon and the interface. (d--f) Similar results for a setup with two interfaces.  $B_0\approx910$~T and $\Delta=0$ in all panels.}~\label{Fig:4}
\end{figure}
%%%%%%%
%%-------------------- end Fig.4--------------------

%==========================================

\emph{\textbf{Device applications---}}
Given the robustness of these edge states, it is natural to look for potential device applications. A realistic proposal is shown in Fig.~\ref{Fig:3}(a), where a ZZ-terminated graphene ribbon is deposited on a hBN substrate with the corrugation $h(\bl{r})$ described above. This setup is acted upon top and bottom gates that induce a scalar potential with the same profile of the corrugation due to the membrane height's variation. Hence, the resulting structure is a realization of the gapped---due to the presence of the hBN substrate---continuum model for electron dynamics governed by a pseudomagnetic field and a scalar potential with the same periodicity [see Eq.~(\ref{Eq:Hamiltonian})]. The tuning of gate voltage allows exploring different transport regimes as the band structure evolves at the Fermi level [Fig.~\ref{Fig:3}(b)]: at zero voltage, the system behaves as a normal semiconductor because of the gap ($2\Delta$) induced by the hBN substrate; while at finite voltages, the gap gradually closes and reopens accompanied by the emergence of propagating edge states. This transition can be understood as a consequence of the competition between the effective staggered potential $\Delta_{\rm eff}\sigma_z$ induced by $V(\bl{r})$ and the staggered potential $\Delta\sigma_z$ introduced by the hBN substrate [recall that $\Delta_{\rm eff}<0$ for $V_0>0$, cf. Figs.~\ref{Fig:1}(b, c)]. The band topology is also tuned in this process: At $V_0=0$, the first conduction bands in the $\pm K$ valleys have Chern numbers $C_{c1}^{\pm K}=\mp1$; while the first valence bands are topologically trivial. After the gap closes, the conduction band becomes trivial and the band topology is transferred to the valence bands. 

The electric tunability of the effective staggered potential $\Delta_{\rm eff}\sigma_z$ also offers the opportunity to generate topological in-gap propagating states with appropriate modifications in the device's design. Fig.~\ref{Fig:4}(a) illustrates a split-gate device, where top and bottom gates are partitioned and deposited in separated regions in the sample. The emergence of topological in-gap states at the interface of two regions with opposite voltage settings [Figs.~\ref{Fig:4}(b, c)] can be understood in terms of perturbation theory [cf. Eq.~(\ref{Eq:Heff})], where the effective staggered term $\Delta_{\rm eff}\sigma_z$ has a magnitude/sign that depends on the intensity/direction of the displacement field. The opposite $\Delta_{\rm eff}$ in the two regions of inverted displacement fields signal band-crossing at the interface, with the emergence of propagating interfacial states. The topological nature of these interfacial states is characterized by the quantized topological charge $\pm1$ in the two valleys, which equals the difference of Chern numbers in the two regions ($C_{\rm reg\,1}^{\tau K}-C_{\rm reg\,2}^{\tau K}$, see SI Sec.~S1~G)~\cite{ValleyBulkEdgeMorpurgoPRB2010}.
Similar interfacial states of the same origin also exist in bilayer graphene~\cite{DomainWallMartinPRL2008,DomainWallBilayerGraphenePRL2008,DomainWallJungJeilPRB2011,DomainWallZhangFanPNAS2013,DomainWallPRX2013,DomainWallFengWangNature2015,DomainWallJunZhuNatNanotech2016,DomainWallJunZhuScience2018,DomainWallJunZhuScience2024} and artificial lattices~\cite{KinkStatePhononics,KinkStatePhotonics}.
By depositing electric gates at selected positions, one can produce topologically conducting channels at desired locations in the sample [Fig.~\ref{Fig:4}(b)]. Importantly, the interface can have generic graphene crystalline configurations beyond those in Table~\ref{Table:BoundaryConfiguration&EdgeState}.
These interfacial states can coexist with the edge states in the zero-energy gap present at the physical boundaries of the ribbon [Fig.~\ref{Fig:4}(b) and Fig.~\ref{Fig:4}(c) inset], and they will not hybridize provided the interface is placed far enough to avoid wavefunction overlaps.

By placing the interface away from the sample boundaries with wide enough gates (e.g., micrometer size), these two types of propagating edge states will remain robust and can be spatially resolved.
The split-gate approach can be extended to include several interfaces with opposite gate voltages positioned at various locations in the sample [Fig.~\ref{Fig:4}(d--f)]. These interfacial states are new conducting channels that can be turned on and off by simply switching the gate voltages in the corresponding regions.

In summary, we have demonstrated that periodically strained graphene superlattices provide a versatile platform for engineering robust electronic states and controlling nanoscale transport. The interplay between strain-induced pseudomagnetic fields and electric displacement-field-generated scalar potentials gives rise to isolated narrow bands with nontrivial valley topology and, remarkably, to propagating edge states within the zero-energy gap despite a vanishing total Chern number. We showed that these edge states originate from modulation of the boundary onsite potential by the periodic scalar potential, revealing a mechanism distinct from conventional topological protection. Although their existence depends on nanoribbon termination, the states exhibit striking robustness against boundary disorders, maintaining nearly quantized conductance under realistic perturbations. Building on these findings, we proposed device architectures in which low-energy edge channels can be electrically switched on and off through competition between corrugation-induced and substrate-induced staggered potentials. Furthermore, split-gate configurations enable the creation of topologically protected interfacial channels, whose location and connectivity can be dynamically controlled by gate voltages. Together, these results establish corrugated graphene superlattices as a promising platform for electrically programmable transport, reconfigurable current pathways, and future straintronic and quantum electronic devices.

\noindent\textbf{Supporting Information---} 
Details of tight-binding calculations, scaling behavior of bulk energy bands, origin of edge states in the zero-energy gap, role of pseudomagnetic field on edge states in high-energy gaps, effects of edge vacancies, comparison between real and pseudo magnetic fields, valley-resolved Chern numbers in the interfacial structure, results for two alternative superlattice models.

\emph{\textbf{Acknowledgment---}}
A.B. acknowledges computing time on the supercomputers at the Center for Research in Molecular Modeling (CERMM), Richard J. Renaud Science Complex, Concordia University.

%\bibliography{RibbonsRefs}

% use the following to remove "References" section in TOC of Supplemental Materials

\let\oldaddcontentsline\addcontentsline% Store \addcontentsline
\renewcommand{\addcontentsline}[3]{}% Make \addcontentsline a no-op
\bibliography{RibbonsRefs}
\let\addcontentsline\oldaddcontentsline% Restore \addcontentsline

\clearpage 
%==========================================

%=====================================================
%  Supplemental Materials

\clearpage % to start with a new page

% To number supplemental material with 'S':
\renewcommand{\thefigure}{S\arabic{figure}}       % \Roman{}, \arabic{}, \Alph{}
\renewcommand{\thesection}{S\arabic{section}}
\renewcommand{\thepage}{S\arabic{page}}
\renewcommand{\theequation}{S\arabic{equation}}
\renewcommand{\thetable}{S\arabic{table}}
\setcounter{figure}{0}
\setcounter{section}{0}
\setcounter{page}{1}
\setcounter{equation}{0}
\setcounter{table}{0}

\title{Supporting Information for ``Propagating edge and interfacial states in corrugated graphene: Robustness and configurability''}
\maketitle
\onecolumngrid
\tableofcontents

%===========================================
\clearpage

\section{Implementation of strain superlattice: model I}

Here we refer to the model described in the main text as model I:
\begin{equation}
	\left.
	\begin{aligned}
		B(\bl{r})&=B_0\sum_{i=1}^{3}\cos(\bl{b}_i\cdot\bl{r})\\
		V(\bl{r})&=V_0\sum_{i=1}^{3}\cos(\bl{b}_i\cdot\bl{r})
	\end{aligned}
	\right\}
	\text{ Model I}
\end{equation}
In the following, we discuss its tight-binding implementation and provide additional results and discuss several properties.

\subsection{Details of the tight-binding calculations}

\begin{figure}[t]
	\centering
	\includegraphics[width=3in]{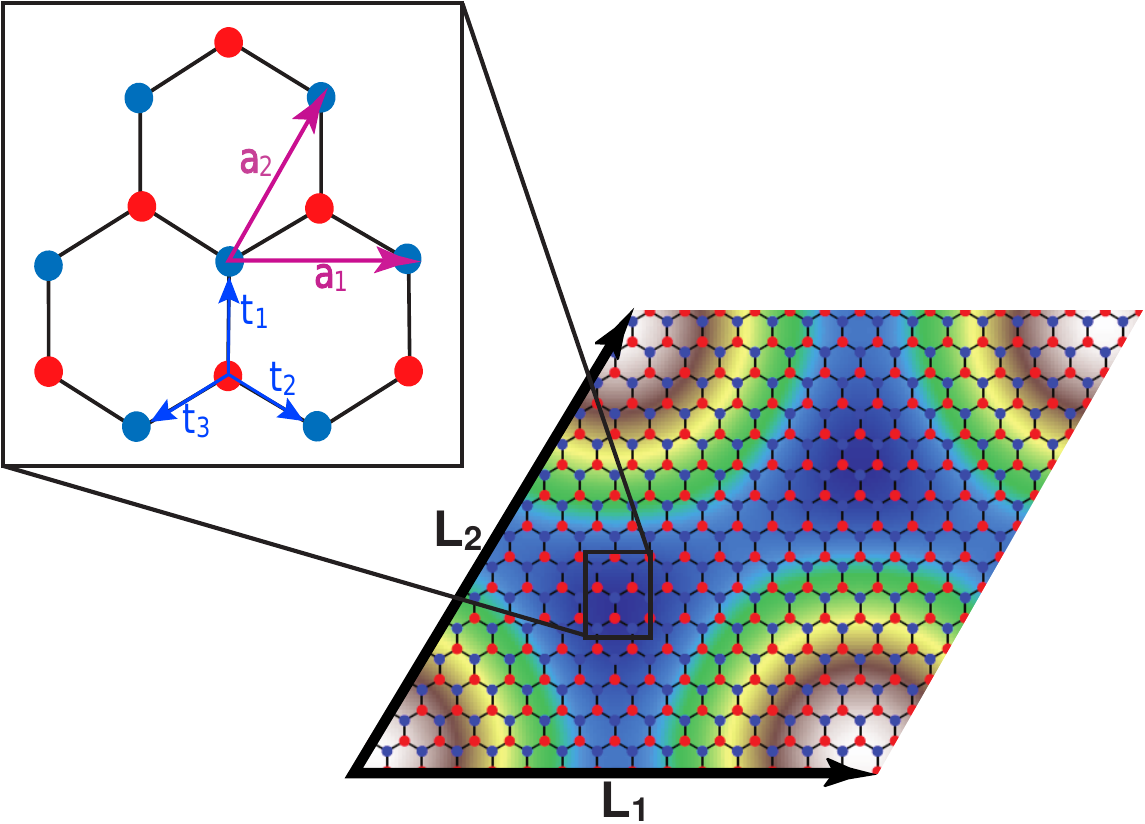}
	\caption{Schematics of primitive lattice vectors of pristine graphene and the strain superlattice.}~\label{Fig:S1}
\end{figure}

We adopt the following general form for the tight-binding model:
%%%-------------- Eq.A. 1 -------------------------
\begin{equation}
	H = -\sum_{\langle{i,j}\rangle} t_{ij}(\bl{r}) \, c_i^{\dagger} c_j+\sum_{i} V(\bl{r}_i) c_i^{\dagger} c_i+\Delta\sum_{i\in A}c_i^{\dagger}c_i-\Delta\sum_{i\in B}c_i^{\dagger}c_i,~\label{eqA1}
\end{equation}
where $c_i^{\dagger}$ ($c_i$) is the creation (annihilation) operator for an electron at site $i$, $t_{ij}(\bl{r})$ is the hopping energy between nearest-neighbor sites $i$ and $j$, $V(\bl{r})$ is the periodic scalar potential, and $\Delta$ is the staggered potential of opposite signs in A and B sublattices. In this work, the numerical results are obtained by using the Kwant package~\cite{Groth2014Kwant}. To this end, we define our primitive lattice vectors of graphene as
%%%-------------- Eq.A. 2 -------------------------
\begin{equation}
	\bl{a}_1= a\left( 1, \ 0 \right) \ \text{and} \,\,\,\, \bl{a}_2 =a\left(\frac{1}{2}, \ \frac{\sqrt{3}}{2} \right),~\label{eqA2}
\end{equation}
where $a$ is the lattice constant of graphene. 
The primitive lattice vectors of the strained superlattice are then obtained as $\bl{L}_i=N\bl{a}_i$, where $N$ is an integer that defines the size of the superlattice unit cell (Fig.~\ref{Fig:S1}). In this work, we use $N=21$. The primitive reciprocal lattice vectors of the superlattice are 
%%%-------------- Eq.A. 3 -------------------------
% \begin{equation}
% \bl{G}_1= \frac{4\pi}{\sqrt{3}L}\left(\frac{\sqrt{3}}{2}, \ -\frac{1}{2} \right) \ \text{and}\ \;\;\bl{G}_2 = \frac{4\pi}{\sqrt{3}L}\left(0, \ 1 \right),~\label{eqA3}
% \end{equation}
\begin{equation}
	\bl{b}_1= \frac{4\pi}{\sqrt{3}L}\left(0,\,1\right)\,\text{ and }\,\bl{b}_2 = \frac{4\pi}{\sqrt{3}L}\left(-\frac{\sqrt{3}}{2}, \ -\frac{1}{2} \right),~\label{eqA3}
\end{equation}
where $L=Na$ denotes the length of the superlattice primitive vectors.

Following Ref.~\cite{PhongPRL2022}, we now provide details of $t_{ij}(\bl{r})$ and $V(\bl{r})$ in Eq.~(\ref{eqA1}). The former is associated with the pseudomagnetic field $B(\bl{r})$. In this work we choose the profile of the periodic pseudomagnetic field (in $\bl{K}$ valley) and scalar potential as~\cite{JinhaiNature2020}
%%%-------------- Eq.A. 4 -------------------------
\begin{equation}
	\bl{B}(\bl{r})=B_0\hat{z}\sum_{i=1}^{3}\cos({\bl{b}}_i \cdot {\bl{r}}) ~~\text{ and }~~ V(\bl{r})=V_0\sum_{i=1}^{3}\cos({\bl{b}}_i\cdot{\bl{r}}),~\label{eqA4}
\end{equation}
where $B_0$ and $V_0$ denote their strength and $\bl{b}_3=-\bl{b}_1 -\bl{b}_2$.
In the presence of strain, the hopping energy is modified as $t_{ij}(\bl{r})=t_0e^{-\beta\left( \frac{r_{ij}}{a_{0}}-1 \right)}$, where $t_0=2.7$~eV and $a_0=0.142$~nm are hopping energy and carbon-carbon distance of non-strained graphene, $\beta=3.37$, and $r_{ij}=\left| \bl{r}_i - \bl{r}_j \right|$~\cite{Ando2005}. The vector potential $\bl{A}(\bl{r})=(A_x,\,A_y)$ corresponding to $\bl{B}(\bl{r})$, i.e., $\bl{B}= \nabla \times \bl{A}$, is connected to the change of hopping energy due to strain~\cite{PhongPRL2022}:
\begin{equation}
	\begin{aligned}
		A_x(\boldsymbol{r}_i) &= -\frac{1}{e v} \left[ \delta t_1(\boldsymbol{r}_i) - \frac{1}{2} \delta t_2(\boldsymbol{r}_i) - \frac{1}{2} \delta t_3(\boldsymbol{r}_i) \right] \\
		A_y(\boldsymbol{r}_i) &= -\frac{1}{e v} \left[ \frac{\sqrt{3}}{2} \delta t_2(\boldsymbol{r}_i) - \frac{\sqrt{3}}{2} \delta t_3(\boldsymbol{r}_i) \right]
	\end{aligned},
\end{equation}
where $\delta t_{j}(\bl{r}_i)= t_{ij}-t_0$ and $v=\frac{\sqrt{3}at_0}{2\hbar}$ is the Fermi velocity.
With these definitions, the hopping $t_{ij}$ in Eq.~(\ref{eqA1}) is straightforwardly given by $t_{ij}=t_0+\delta t_j (\bl{r}_i)$, where
%%%-------------- Eq.A. 7 -------------------------
\begin{equation}
	\delta t_j (\bl{r}_i)= \frac{\sqrt{3}evB_0L}{4\pi}\sin(\bl{b}_j\cdot\bl{r}_i)\approx\frac{B_0}{9.1\times10^4~\text{T}}Nt_0\sin(\bl{b}_j\cdot\bl{r}_i).~\label{eqA7}
\end{equation}

For this model, we use a ribbon of width equals to $6.1L=31.51$~nm.

\subsection{Scaling behavior of the bulk bands}~\label{Sec.ScalingTriangularB}

Now we show that the bulk energy bands exhibit some scaling behavior that is determined by the factor $\Phi\propto B_0L^2$. We illustrate this by using the continuum model in the $K$ valley: $\mathcal{H}_+=v\boldsymbol{\sigma}_+\cdot\left(\bl{p}+e\bl{A}\right)$, where only the vector potential is considered at the moment. 
The pseudomagnetic field profile $B(\bl{r})=B_0\sum_{i=1}^{3}\cos(\bl{b}_i\cdot\bl{r})$ can be achieved with the vector potential $\bl{A}=(A_x,\,0)$, where
\begin{equation}
	eA_x
	=\text{sgn}(B_0)\frac{\hbar}{l_0^2}\frac{\sqrt{3}L}{2\pi}\left[-\frac{1}{2}\sin(\bl{b}_{1}\cdot\bl{r})+\sin(\bl{b}_{2}\cdot\bl{r})+\sin(\bl{b}_{3}\cdot\bl{r})\right].
\end{equation}
In the above, we have defined the magnetic length $l_0=\sqrt{\frac{\hbar}{e|B_0|}}$.
The pristine part of the Hamiltonian can be rewritten as
\begin{equation}
	H_0=v\boldsymbol{\sigma}_+\cdot\bl{p}=E_0\boldsymbol{\sigma}_+\cdot\tilde{\bl{k}},
\end{equation}
where $E_0=\hbar v|\bl{b}_i|=\hbar v\frac{4\pi}{\sqrt{3}L}$ characterizes the strength of the pristine kinetic energy and $\tilde{\bl{k}}=\bl{k}/|\bl{b}_i|$ is dimensionless. For the continuum model, we use $\hbar v \approx 6.582$~eV$\cdot$\AA~throughtout the manuscript. The potential term of the Hamiltonian can be rewritten as
\begin{equation}
	H_v=ev \boldsymbol{\sigma}_+\cdot\bl{A}=\frac{\hbar v L}{l_0^2}\sigma_x \tilde{A}_x,
\end{equation}
where $\tilde{A}_x=\text{sgn}(B_0)\frac{\sqrt{3}}{2\pi}\left[-\frac{1}{2}\sin(\bl{b}_{1}\cdot\bl{r})+\sin(\bl{b}_{2}\cdot\bl{r})+\sin(\bl{b}_{3}\cdot\bl{r})\right]$. 
In these forms, the Hamiltonian becomes
\begin{equation}
	\mathcal{H}_+=E_0\boldsymbol{\sigma}_+\cdot\left(\tilde{\bl{k}}+\frac{\hbar v L}{l_0^2E_0}(\tilde{A}_x,\,0)\right)=E_0\boldsymbol{\sigma}_+\cdot\left(\tilde{\bl{k}}+\Phi(\tilde{A}_x,\,0)\right),
\end{equation}
where $\Phi=\frac{\hbar v L}{l_0^2E_0}\propto B_0L^2$ characterizes the ratio of the vector potential energy to the kinetic energy and is proportional to the magnetic flux in the unit cell. The eigenenergy problem of solving $\mathcal{H}_+\psi=E\psi$ is equivalent to
\begin{equation}
	\boldsymbol{\sigma}_+\cdot\left(\tilde{\bl{k}}+\Phi(\tilde{A}_x,\,0)\right)\psi=\tilde{E}\psi,
\end{equation}
where $\tilde{E}=E/E_0$.
From this, it becomes clear that if we measure the wave vector in units of $|\bl{b}_i|$, and energy in units of $E_0$, the factor $\Phi\propto B_0L^2$ is the only independent tuning parameter in the model. Generally, a larger $\Phi$ leads to narrower bands. 

In the presence of a small periodic scalar potential and/or constant staggered potential, as discussed in the main text, a gap opens at the Dirac point, while the dispersions at higher energies are modestly affected.

\begin{figure}[h!]
	\centering
	\includegraphics[width=3.7in]{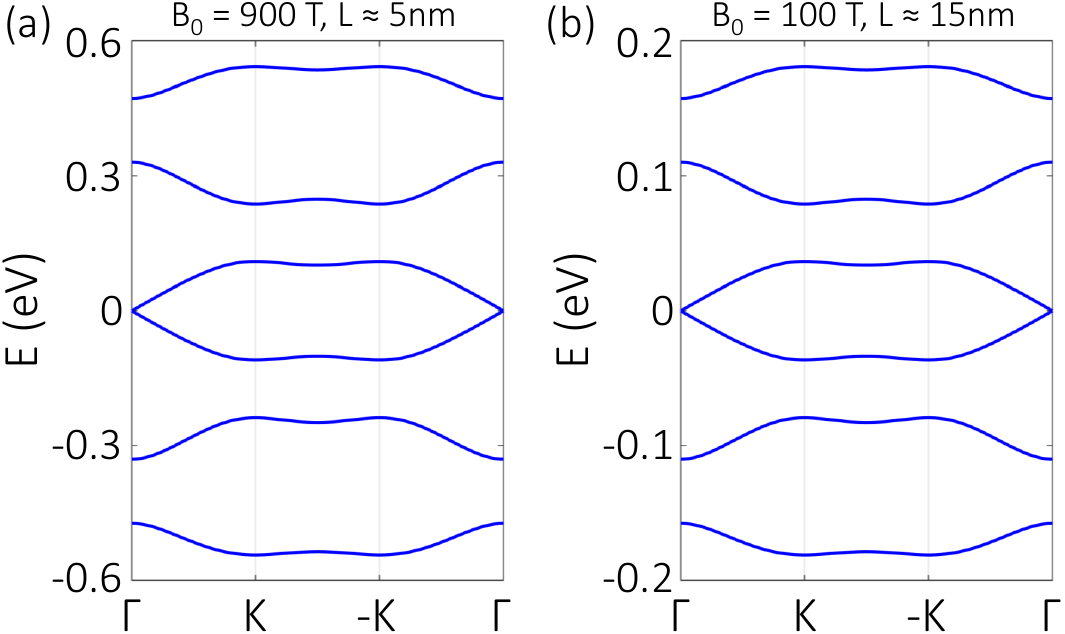}
	\caption{(a) $K$ valley continuum-model energy bands with $B_0=900$~T and $L\approx5$~nm. (b) Similar results as in (a) but with $B_0=100$~T and $L\approx15$~nm. $V_0=\Delta=0$ in both panels.}~\label{Fig:PhongModel_ContinuumBands}
\end{figure}

Fig.~\ref{Fig:PhongModel_ContinuumBands} illustrates the scaling behavior of the bulk energy bands explicitly using two superlattice configurations of identical $\Phi$, where it is clear that the bands in panels (a) and (b) are identical if measured in unis of $E_0\propto1/L$. As the values of $B_0$ and $L$ for Fig.~\ref{Fig:PhongModel_ContinuumBands}(a) are similar to those for Fig.~1(b) of the main text, the bands in Fig.~\ref{Fig:PhongModel_ContinuumBands}(a) are consistent with those in Fig.~1(b) up to the absence of a zero-energy gap.

In our tight-binding calculations, a small period $L$ is considered to reduce computational cost, thus an artificially large $B_0$ is used to flatten the bands. In experimentally realistic setups, the period can be of the order of $L\sim\mathcal{O}(10)$~nm, a much smaller $B_0$ is enough to generate narrow bands [e.g., compare Figs.~\ref{Fig:PhongModel_ContinuumBands}(a) and (b)].

\subsection{Origin of propagating edge states in the zero-energy gap}~\label{Sec:SM_EdgeStateOrigin}

\begin{figure}[h!]
	\centering
	\includegraphics[width=6in]{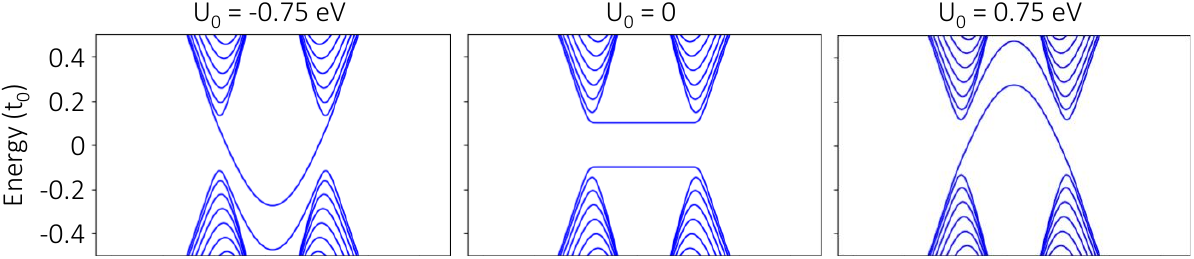}
	\caption{Energy bands of pristine zigzag graphene nanoribbon with a staggered potential $\Delta\sigma_z$ and onsite potential $U_0$ on its two boundaries. $\Delta=0.1 t_0$ in all the panels.}~\label{Fig:S_PristineGraphene}
\end{figure}

To understand the presence of edge states in the zero-energy gap, it is intuitive to first consider the scenario of a periodically strained graphene ribbon with a staggered potential $\Delta<0$ [Fig.~1(b)], where the conduction and valence band edges exhibit exact flat bands at energy $\pm\Delta$. The states of the conduction (valence) flat band are $\mathcal{B}$ ($\mathcal{A}$) sublattice polarized and localized on the lower (upper) boundary of the ribbon [cf. Fig.~1(a)].
Similar degenerate flat bands also exist in a pristine zigzag graphene nanoribbon, and become dispersive in the presence of an onsite potential $U_0$ added at the boundary, developing positive or negative velocities and crossing the bandgap for appropriate potential intensities (see e.g., Fig.~\ref{Fig:S_PristineGraphene})~\cite{YaoEdgeStates2009}.
In the case of Fig.~1(c), the pseudomagnetic field and the scalar potential lead to an effective staggered potential $\Delta_{\rm eff}<0$. Naively, one might expect two flat bands similar to those in Fig.~1(b).
Notice, however, that the scalar potential plays an additional role: it modulates the energy on the ribbon boundaries. In the configuration of Fig.~1(a), the potential $V(\bl{r})$ on the boundaries is negative,  pulling the flat conduction band down and leading to the in-gap propagating dispersion. The corresponding edge states are $\mathcal{B}$ sublattice polarized and reside at the lower boundary [Fig.~1(c)]. Interestingly, some flat valence bands remain in the spectrum, whose states reside on the upper boundary where $V(\bl{r})$ has its minima [Fig.~1(a)].

%==========================================================

\subsection{Pseudomagnetic field-induced sublattice polarization and its connection to the boundary location of edge states in higher energy gaps}

\begin{figure}[h!]
	\centering
	\includegraphics[width=6in]{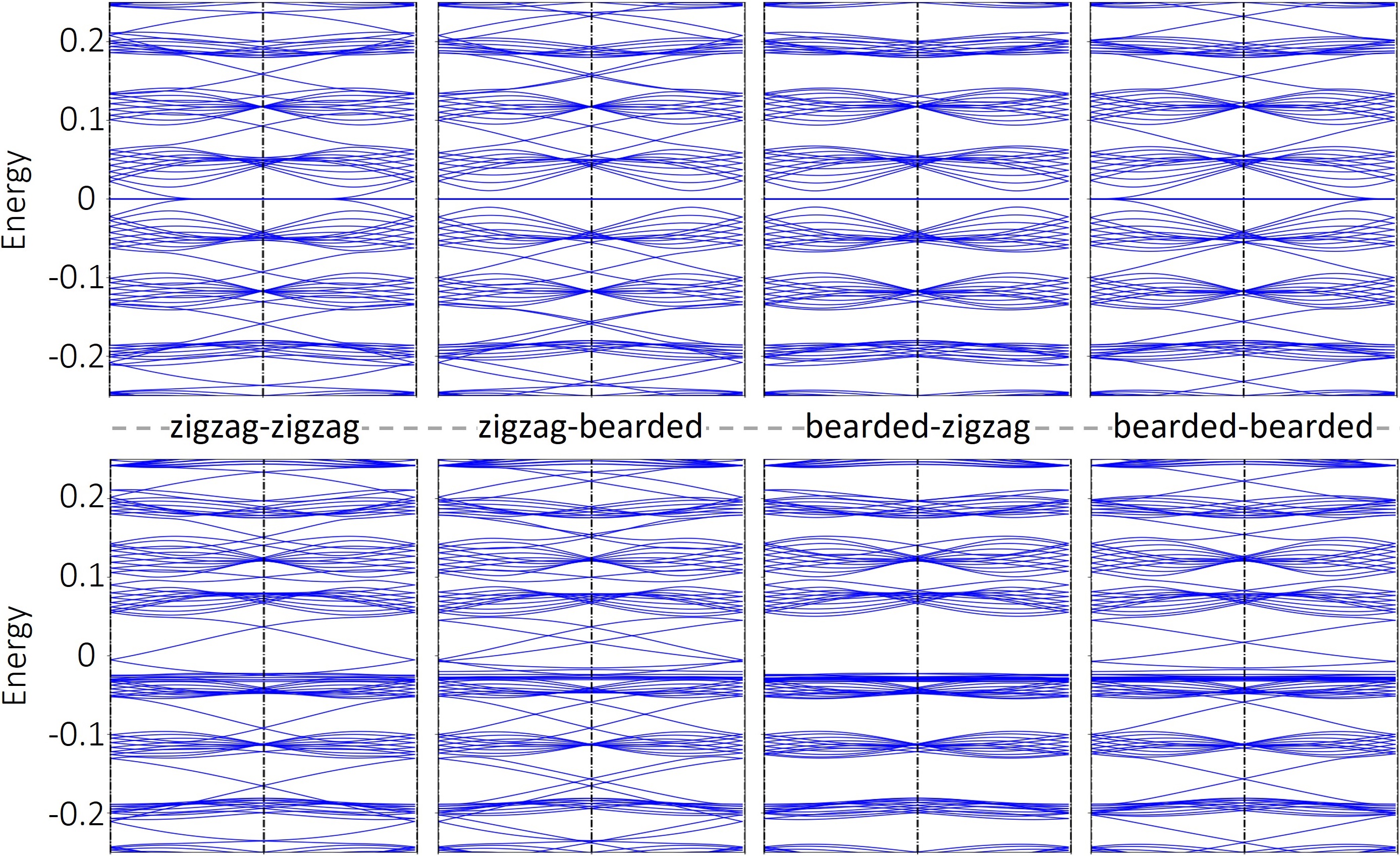}
	\caption{\textbf{Edge states in a graphene nanoribbon with different (bottom-top) boundary configurations}. Energy bands of the periodically strained nanoribbon with a pseudomagnetic field $B(\bl{r})=B_0h(\bl{r})$ in the absence (first row) and presence (second row) of a periodic scalar potential $V(\bl{r})=V_0h(\bl{r})$. $B_0\approx1365$~T in all panels, $V_0=30$~meV in the second row.}~\label{Fig:DifferentEdge}
\end{figure}

In the case of uniform pseudomagnetic fields, the in-gap propagating edge states in periodically strained graphene were found to originate from the hybridization of the sublattice-polarized edge states of a pristine graphene nanoribbon with the sublattice-polarized pseudo-Landau level-like bulk states in the presence of a pseudomagnetic field~\cite{EdgeStateLocationPRB2017}. The edge state disperses across the gap when the hybridization occurs between the two types of states having opposite sublattice polarizations. In our case, where the pseudomagnetic field is non-uniform with vanishing flux per unit cell [Fig.~1(a)], we found that the $\varhexagon$ part of the pseudomagnetic field is most crucial. Explicit numerical results show that in the $\varhexagon$ region where the pseudomagnetic field is negative in the $K$ valley [the blue area in Fig.~1(a)], the bulk states are $\mathcal{A}$ sublattice polarized. Consequently, the in-gap propagating edge states will appear on the boundary composed of the $\mathcal{B}$ sublattices: lower boundary if it has the zigzag configuration and/or upper boundary if it has bearded configuration (Table~I of the main text). The presence or absence of edge states (except in the zero-energy gap) in Fig.~\ref{Fig:DifferentEdge} is consistent with this.

%%---------------------- Fig.S1 --------------------
%%%%%
%%%%%%%
%%-------------------- end Fig.S1--------------------

\subsection{Effects of edge vacancies/defects on the edge states}

To investigate the robustness of the edge states against boundary imperfections, we introduce vacancy disorder that is spatially confined to the zigzag edges of the graphene ribbon. The width of the disordered region, $W_{\mathrm{edge}}$, is determined from the spatial extent of the localized edge states obtained in the absence of disorder [cf. Fig.~1(d) of the main text]. Vacancies are randomly introduced within this region with a concentration $V_{\%}$, expressed as a percentage of the total number of lattice sites within the corresponding edge region, while the central bulk region remains completely disorder-free. Thus, for example, $V_{\%}=5\%$ corresponds to the random removal of $ 5\%  $ of the lattice sites within the edge-disordered region.

The vacancy disorder can formally be described by
\begin{equation}
	H_{\mathrm{dis}}= V_{\mathrm{vac}} \sum_{i\in\mathcal{V}_{\mathrm{edge}}} c_i^\dagger c_i
	\quad\text{with}\quad
	V_{\mathrm{vac}}\rightarrow\infty,
\end{equation}
where $\mathcal{V}_{\mathrm{edge}}$ denotes the set of randomly selected vacancy sites within the edge-disordered region. In the numerical calculations, vacancies are implemented by directly removing the selected sites from the tight-binding lattice, rather than by assigning a large onsite potential ($V_{\mathrm{vac}}\rightarrow\infty$) to these sites. Consequently, both the corresponding orbitals and all hopping matrix elements connected to the removed sites are eliminated from the Hamiltonian. The number of removed sites is determined by the vacancy concentration $V_{\%}$. 
%This procedure ensures that the disorder is strictly confined to the region associated with the localized edge states, leaving the bulk of the graphene ribbon unaffected.

\begin{figure}[t]
	\centering
	\includegraphics[width=5in]{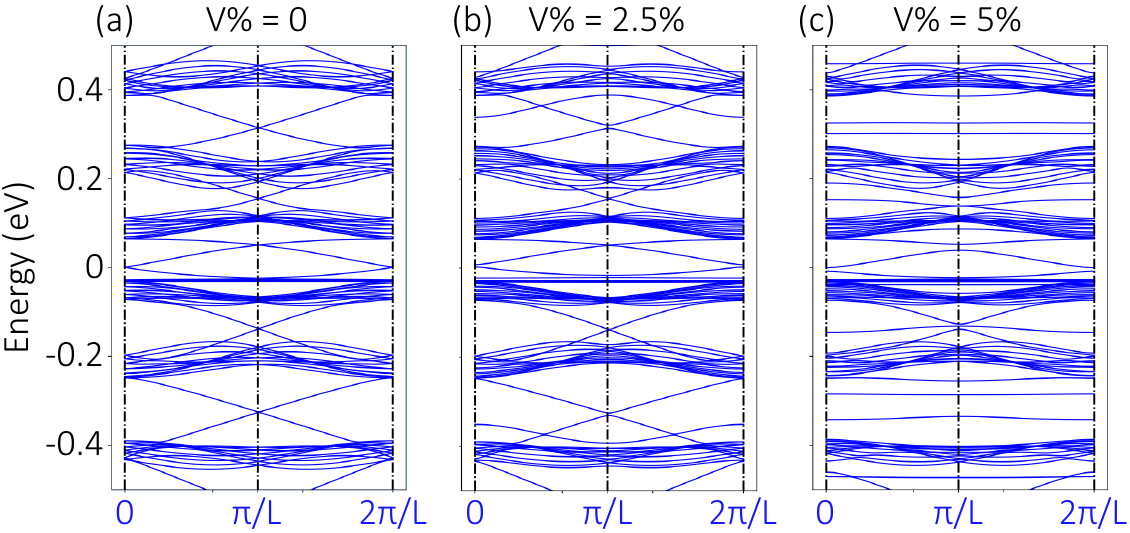}
	\caption{Effects of vacancy percentage on edge state dispersion. $V_{\%}=$ (a) 0\%, (b) 2.5\%, (c) 5\%. $B_0\approx910$~T, $V_0=30$~meV and $\Delta=0$ in all panels.}~\label{Fig:S_vacancy}
\end{figure}

Fig.~\ref{Fig:S_vacancy} illustrates the effects of boundary vacancies on the edge state dispersions. When the vacancy percentage $V_{\%}$ is moderate [e.g., Fig.~\ref{Fig:S_vacancy}(b)], the edge state dispersions are nearly unaffected with tiny gap openings due to inter-valley scattering. One observes that some of the high-energy bulk bands are shifted into the neighboring gaps and intersect with the edge state dispersions. As the vacancy percentage $V_{\%}$ becomes larger [e.g., Fig.~\ref{Fig:S_vacancy}(c)], larger gap openings emerge in the edge state dispersions. Interestingly, in this case the edge states in the zero-energy gap appear to be more robust (than the higher-energy ones), for which the majority of the dispersion remains intact.

%===========================================
\subsection{Comparison between real and pseudomagnetic fields}

In the case of a real magnetic field, the continuum model in the ($\mathcal{A},\,\mathcal{B}$) sublattice basis becomes
\begin{equation}
	H_{\rm r}^{\tau}=v\bl{\sigma}_{\tau}\cdot(\bl{p}+e\bl{A})+V(\bl{r})+\Delta\sigma_z,
\end{equation}
where both valleys have the same vector potential $\bl{A}$ satisfying $\nabla\times\bl{A}(\bl{r})=\bl{B}(\bl{r})$, and all the other quantities are the same as those defined in the main text. 
In the tight-binding formalism, the presence of a real magnetic field is implemented by using Peierls substitution. In this case, the hopping energy $ t_{ij}$, in Eq.~(\ref{eqA1}), is expressed as
%%%-------------- Eq.A. 8 -------------------------
\begin{equation}
	t_{ij}=t_0 e^{-i\frac{e}{\hbar}\int_{r_i}^{r_j} \bl{A}(\bl{r})\cdot d{\bl{r}}}.~\label{eqA8}
\end{equation}
For the magnetic field $\bl{B}(\bl{r})=B_0\hat{z}\sum_{i=1}^{3}\cos({\bl{G}}_i \cdot {\bl{r}})$, the vector potential $\bl{A}({\bf r})$ can be chosen in components as
%%%-------------- Eq.A. 9 -------------------------
\begin{equation}
	\begin{aligned}
		& A_x(\bl{r})=-\frac{\sqrt{3}B_0 L}{4 \pi}\left[\sin \left(\bl{G}_1 \cdot \bl{r}\right)-\frac{1}{2} \sin \left(\bl{G}_2 \cdot \bl{r}\right)-\frac{1}{2} \sin \left(\bl{G}_3 \cdot \bl{r}\right)\right]\\
		& A_y(\bl{r})=-\frac{\sqrt{3}B_0 L}{4 \pi}\left[\frac{\sqrt{3}}{2} \sin \left(\bl{G}_2 \cdot \bl{r}\right)-\frac{\sqrt{3}}{2} \sin \left(\bl{G}_3 \cdot \bl{r}\right)\right]
	\end{aligned}.~\label{eqA9}
\end{equation}

\begin{figure*}[h!]
	\centering
	\includegraphics[width=6in]{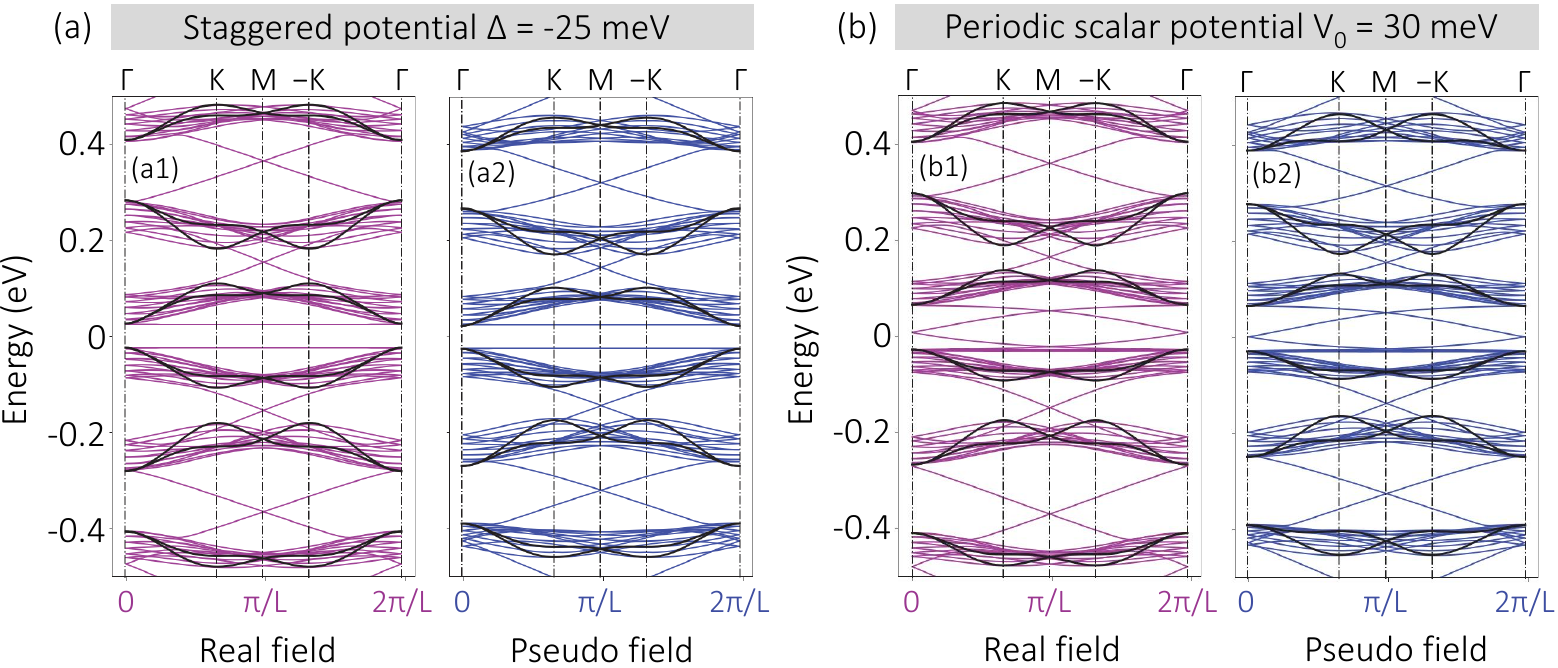}
	\caption{\textbf{Edge states in a zigzag-zigzag nanoribbon: comparison between real and pseudo magnetic field scenarios}. (a) Energy bands of the nanoribbon with a staggered potential $\Delta \sigma_z$ in a real \textit{vs} pseudo magnetic field $B(\bl{r})=B_0h(\bl{r})$. The black curves represent bulk energy bands with high symmetry points labeled on the top. (b) Similar to those in panel (a), but with the staggered potential replaced by a periodic scalar potential $V(\bl{r})=V_0h(\bl{r})$. $B_0\approx910$~T in all panels.}~\label{FigS_real_vs_pseudo}
\end{figure*}

Fig.~\ref{FigS_real_vs_pseudo} shows the energy bands of a zigzag-zigzag graphene nanoribbon in the presence of a real or pseudo magnetic field. Panel (a) illustrates the comparison when a staggered potential is included, while panel (b) corresponds to the comparison with a periodic scalar potential $V(\bl{r})$. The energy bands are nearly identical for a real vs pseudo magnetic field. The Chern numbers of the bands, however,
are very different, as summarized in Table~\ref{Table:ChernNumber}. 
Additionally, one finds identical (different) Chern numbers for the pseudo (real) field cases in Figs.~\ref{FigS_real_vs_pseudo}(a2, b2) (Figs.~\ref{FigS_real_vs_pseudo}(a1, b1)).
These variations in the Chern numbers can be understood from the perturbation theory analysis [cf. Eq.~(2) of the main text]: In the case of a real magnetic field, the scalar potential $V(\bl{r})$ leads to $m_{\rm eff}$ of \textit{opposite} signs in the two valleys, consistent with the breaking of time-reversal symmetry. 

Finally, we remark that the feature that the energy bands for a real vs pseudo magnetic field in Fig.~\ref{FigS_real_vs_pseudo} appear to be (nearly) identical is accidental. The energy bands of the in-gap edge states for a pseudo magnetic field strongly depend on the ribbon's edge configuration (Fig.~\ref{Fig:DifferentEdge}) as they are not topological in nature. While the edge states in a real magnetic field are topological, thus their dispersions (more precisely, the number of edge states) are independent on the edge configuration and determined by the total Chern number $C_{\rm tot}=\sum_{n}(C_{n}^{K}+C_n^{-K})$ of energy bands below the gap.

%The opposite vs identical $m_{\rm eff}$ in the real vs pseudo magnetic field affects the Chern numbers of the energy bands.

%-------------------- table ---------------------------
\begin{table*}[h!]
	\caption{Summary of the Chern number $C_{n}^{\tau K}$ of the first conduction ($n=1$) or valence ($n=-1$) bulk band in the $\tau K$ valley and the total Chern number $C_{\rm tot}=\sum_{n<0}(C_{n}^{K}+C_n^{-K})$ of all the valence bands (denoted as Chern number in the zero-energy gap) for the four cases shown in Figs.~\ref{FigS_real_vs_pseudo}. All the other energy bands shown in Figs.~\ref{FigS_real_vs_pseudo} have zero Chern numbers, and note that there exist other bands with zero or nonzero Chern numbers outside the shown energy window. Whether propagating edge states can exist (\ding{51}) or not (\ding{55}) in the zero-energy gap is also indicated.}~\label{Table:ChernNumber}
	\begin{ruledtabular}
		\begin{tabular}{c|cccc|cccc}
			\multirow{2}{*}{}   & 
			\multicolumn{4}{c|}{Staggered potential $\Delta=-25$ meV}  & 
			\multicolumn{4}{c}{Periodic scalar potential $V_0=30$ meV} \\ \cline{2-9} 
			& \multicolumn{2}{c|}{\textcolor{purple}{Real field (a1)}}                                    & \multicolumn{2}{c|}{\textcolor{RoyalBlue}{Pseudo field (a2)}}            & \multicolumn{2}{c|}{\textcolor{purple}{Real field (b1)}}                                    & \multicolumn{2}{c}{\textcolor{RoyalBlue}{Pseudo field (b2)}}            \\ \hline
			1st conduction band & \multicolumn{1}{c|}{\textcolor{purple}{$C^{K}_{1}=0$}}  & \multicolumn{1}{c|}{\textcolor{purple}{$C^{-K}_{1}=-1$}} & \multicolumn{1}{c|}{\textcolor{RoyalBlue}{$C^{K}_{1}=0$}}  & \textcolor{RoyalBlue}{$C^{-K}_{1}=0$} & \multicolumn{1}{c|}{\textcolor{purple}{$C^{K}_{1}=0$}}  & \multicolumn{1}{c|}{\textcolor{purple}{$C^{-K}_{1}=0$}}  & \multicolumn{1}{c|}{\textcolor{RoyalBlue}{$C^{K}_{1}=0$}}  & \textcolor{RoyalBlue}{$C^{-K}_{1}=0$} \\ \hline
			In zero-energy gap  & \multicolumn{2}{c|}{\textcolor{purple}{$C_{\rm tot}=0$ ~~~\ding{55}}}                               & \multicolumn{2}{c|}{\textcolor{RoyalBlue}{$C_{\rm tot}=0$ ~~~\ding{55}}}         & \multicolumn{2}{c|}{\textcolor{purple}{$C_{\rm tot}=-1$ ~~~\ding{51}}}                              & \multicolumn{2}{c}{\textcolor{RoyalBlue}{$C_{\rm tot}=0$ ~~~\ding{51}}}         \\ \hline
			1st valence band    & \multicolumn{1}{c|}{\textcolor{purple}{$C^{K}_{-1}=-1$}} & \multicolumn{1}{c|}{\textcolor{purple}{$C^{-K}_{-1}=0$}}  & \multicolumn{1}{c|}{\textcolor{RoyalBlue}{$C^{K}_{-1}=-1$}} & \textcolor{RoyalBlue}{$C^{-K}_{-1}=1$} & \multicolumn{1}{c|}{\textcolor{purple}{$C^{K}_{-1}=-1$}} & \multicolumn{1}{c|}{\textcolor{purple}{$C^{-K}_{-1}=-1$}} & \multicolumn{1}{c|}{\textcolor{RoyalBlue}{$C^{K}_{-1}=-1$}} & \textcolor{RoyalBlue}{$C^{-K}_{-1}=1$} \\
		\end{tabular}
	\end{ruledtabular}
\end{table*}
%-------------------- end table ---------------------------

%=========================================================

\subsection{Valley-resolved Chern numbers for $V_0>0$ and $V_0<0$ regions in the interfacial structure}~\label{Sec.ValleyChernInterface}

\begin{figure}[h!]
	\centering
	\includegraphics[width=4.5in]{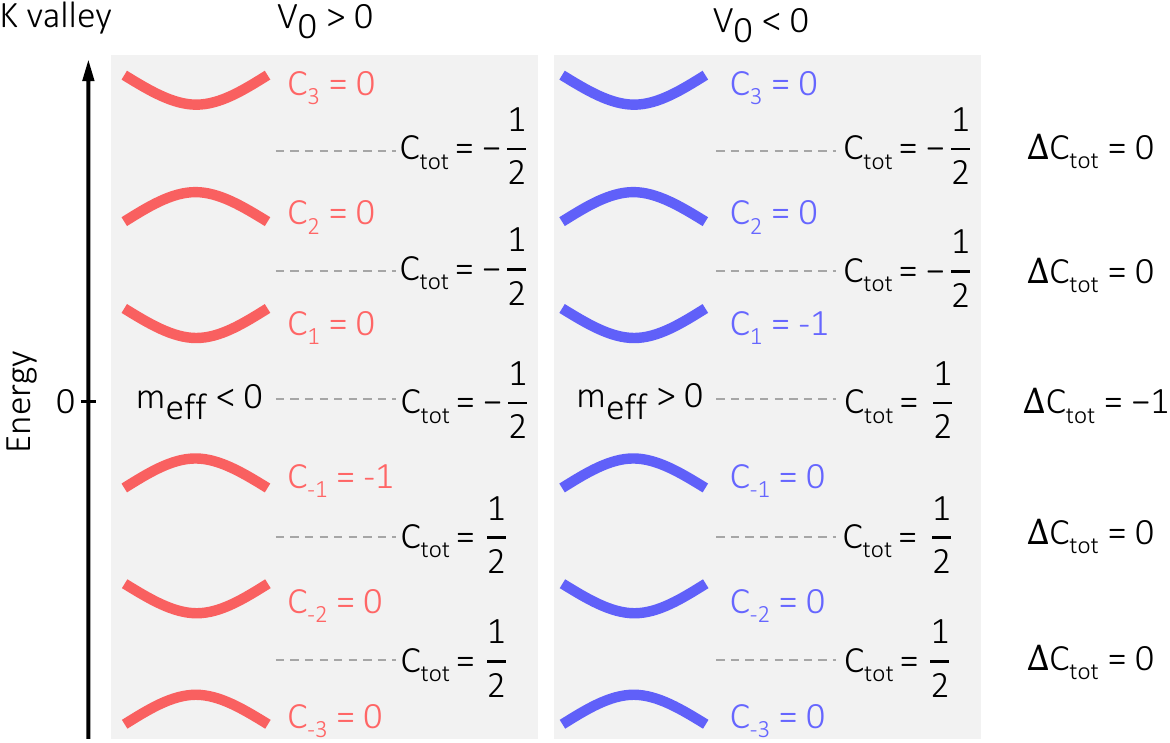}
	\caption{Schematics of bulk energy bands in the K valley for $V_0>0$ and $V_0<0$, where a pseudomagnetic field is also assumed (cf. Fig.~4 of the main text). The corresponding band Chern numbers are shown in red and blue colors. The total Chern number $C_{\rm tot}^{K}=\sum_{n<\text{Gap}}C_{n}^{K}$ of all the bands below a certain gap --denoted by a gray dashed line-- is shown in black. The last column shows $\Delta C^{K}_{\rm tot}=C^{K}_{\rm tot}(V_0>0)-C^{K}_{\rm tot}(V_0<0)$ for different gaps. In the figure for simplicity we have neglected the supersript `K' in the different Chern numbers.}~\label{Fig:S_ValleyChernNumbersInterface}
\end{figure}

Fig.~\ref{Fig:S_ValleyChernNumbersInterface} shows the schematic energy bands in $K$ valley of an interfacial structure discussed in the main text (cf. Fig.~4). The band Chern numbers $C_n^K$ and the total Chern number $C_{\rm tot}^{K}$ of all the energy bands below a certain gap are also shown. The Chern numbers in the $-K$ valley are opposite due to time-reversal symmetry. One finds that the difference in the two regions of opposite scalar potential: $\Delta C_{\rm tot}^{K}=C_{\rm tot}^{K}(V_0>0)-C_{\rm tot}^{K}(V_0<0)=-1$ in the zero-energy gap and vanishes otherwise. Therefore, one pair of interfacial states are expected at each interface with the corresponding dispersion living in the zero-energy gap.

%=========================================================

%NOTE for me: To calculate the Zak phase numerically from a tight-binding band structure, you must compute the Berry phase over the one-dimensional (1D) Brillouin zone. This is done using the Wilson loop method, which involves discretizing the Brillouin zone and calculating overlaps of Bloch wavefunctions. See \href{https://www.google.com/search?client=safari&rls=en&q=how+to+calculate+the+zak+phase+numerically+from+a+band+structure+tight+binding&ie=UTF-8&oe=UTF-8}{this}.

%===========================================

\section{Strain-induced fields by a height profile $z(\bl{r})=z_0\sum_{i=1}^{3}\cos(\bl{b}_i\cdot\bl{r}+\frac{\pi}{4})$ with in-plane displacements: Model I'}~\label{Sec:PhongHeightFile}

\begin{figure}[h!]
	\centering
	\includegraphics[width=7in]{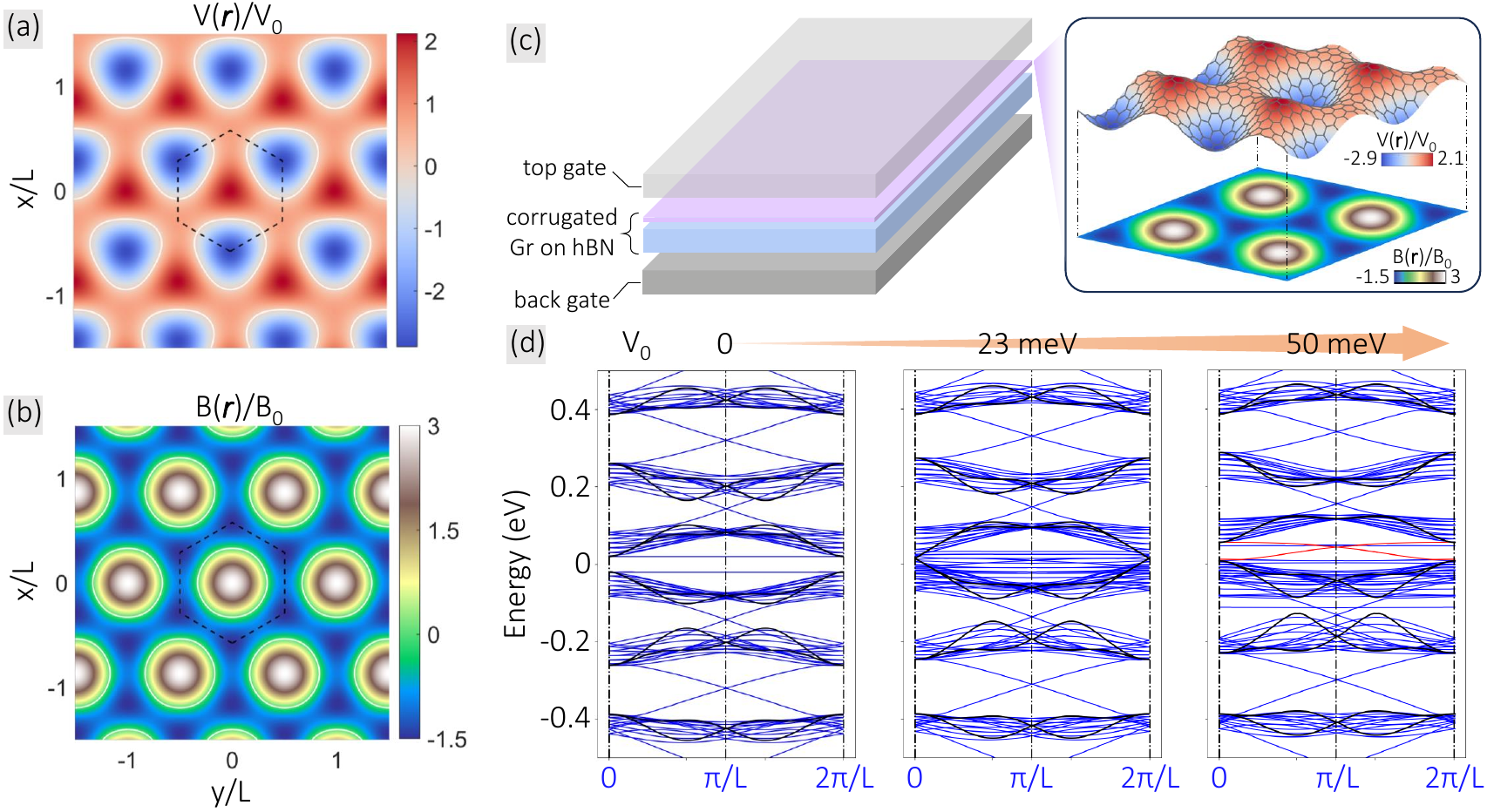}
	\caption{\textbf{Model I'}. (a) Profile of the periodic scalar potential $V(\bl{r})=V_0\sum_{i=1}^{3}\cos(\bl{b}_i\cdot\bl{r}+\frac{\pi}{4})$. (b) Profile of the pseudomagnetic field $B(\bl{r})=B_0\sum_{i=1}^{3}\cos(\bl{b}_i\cdot\bl{r})$. In (a) and (b), the white curves mark contours of $V(\bl{r})=0$ or $B(\bl{r})=0$, the black dashed lines enclose an hexagonal unit cell. (c) Schematics of periodically corrugated graphene on hBN with height profile $z(\bl{r})=z_0\sum_{i=1}^{3}\cos(\bl{b}_i\cdot\bl{r}+\frac{\pi}{4})$. An out-of-plane displacement field can be applied to generate the $V(\bl{r})$ in (a). The hBN introduces a staggered potential $\Delta\sigma_z$ onto graphene. (d) Evolution of the energy bands of a zigzag nanoribbon of width $6.12L=31.61$~nm (blue curves) with respect to $V_0$ by tuning the displacement field. The black curves represent the bulk bands along the same high symmetry path as in Figs.~1(b, c) of the main text. $B_0\approx910$~T and $\Delta=20$~meV have been used.}~\label{Fig:S_model1p}
\end{figure}

A graphene membrane with the height profile 
\begin{equation}
	z(\bl{r})=z_0\sum_{i=1}^{3}\cos(\bl{b}_i\cdot\bl{r}+\frac{\pi}{4})~\label{Eq:PhongHeight}
\end{equation}
was considered in Ref.~\cite{PhongPRL2022}, and the membrane was allowed to relax with in-plane displacements $\bl{u}$ to minimize the elastic energy. 
The resultant strain tensor is given by 
\begin{equation}
	\epsilon_{ij}=\frac{1}2(\partial_i u_j +\partial_j u_i +\partial_i z \, \partial_j z),
\end{equation}
which leads to a pseudovector potential
\begin{equation}
	\bl{A}=-\frac{\sqrt{3}\hbar\beta}{2ae}(\epsilon_{xx}-\epsilon_{yy},\,-2\epsilon_{xy}).
\end{equation}
Here $\beta\approx3$, $a\approx2.46$~\AA\, is lattice constant of graphene, and $\hbar/e\approx65821$  T$\cdot$\AA$^2$ is the (reduced) flux quantum. 
Interestingly, such a relaxed membrane with both in-plane and out-of-plane atomic displacements can produce the pseudomagnetic field profile $\bl{B}(\bl{r})=\nabla\times\bl{A}=B_0\hat{z}\sum_{i=1}^{3}\cos(\bl{b}_i\cdot\bl{r})$ discussed in the main text~\cite{PhongPRL2022}, which has also been confirmed in Ref.~\cite{PeriodicStrainRelaxationPRB2024}. With the height profile of Eq.~(\ref{Eq:PhongHeight}), the scalar potential induced by an out-of-plane electric field reads $V(\bl{r})=V_0\sum_{i=1}^{3}\cos(\bl{b}_i\cdot\bl{r}+\frac{\pi}{4})$, as shown in Fig.~\ref{Fig:S_model1p}(a), which differs from the one discussed in the main text by the $\pi/4$ phase.
As this is close to the model I in main text, here we denote this configuration as model I':
\begin{equation}
	\left.
	\begin{aligned}
		B(\bl{r})&=B_0\sum_{i=1}^{3}\cos(\bl{b}_i\cdot\bl{r})\\
		V(\bl{r})&=V_0\sum_{i=1}^{3}\cos(\bl{b}_i\cdot\bl{r}+\frac{\pi}{4})
	\end{aligned}
	\right\}
	\text{ Model I'}
\end{equation}
This model yields quantitatively similar results as model I in the main text. To highlight this, it suffices to show results analogous to Fig.~3 in the main text. Fig.~\ref{Fig:S_model1p}(d) presents the energy bands of Model I' with $B_0\approx910$~T, $\Delta=20$~meV and three different $V_0$ values. In this model, $B_0>0$ and $V_0>0$ also lead to an effective staggered potential $\Delta_{\rm eff}<0$. As the profiles of $B(\bl{r})$ and $V(\bl{r})$ are different [Fig.~\ref{Fig:S_model1p}(a) vs (b)], the resultant $\Delta_{\rm eff}$ has a smaller magnitude compared to that in model I~\cite{SnymanPRB2009,GrapheneGapTonyLowPRB2011}, thus a larger $V_0$ is needed in Fig.~\ref{Fig:S_model1p}(d) to close the zero-energy gap initially introduced by $\Delta$ [cf. middle panel of Fig.~3(b) of the main text]. Importantly, for a large $V_0$, propagating edge states emerge with energies in the zero-energy gap [Fig.~\ref{Fig:S_model1p}(d) third panel]. 
%We note that localized states characterized by a flat dispersion might coexist with the propagating edge states, but their presence does not affect our discussions of the latter.

%===========================================
\section{Strain-induced fields by a surface superlattice corrugation without in-plane displacements: model II}

In this section, by following Ref.~\cite{TareqNanoLett2023}, we consider a strain superlattice described by out-of-plane surface corrugation 
\begin{equation}
	z(\bl{r})=z_0\sum_{i=1}^{3}\cos(\bl{b}_i\cdot\bl{r}+\phi),
\end{equation}
where $\phi$ is a phase factor that tunes the profile of the corrugation. 
Unlike in Sec.~\ref{Sec:PhongHeightFile}, here we (artificially) turn off in-plane atomic displacements, thus the strain tensor associated with this corrugated profile is
\begin{equation}
	\epsilon_{ij}=\frac{1}{2}\partial_{i}z\,\partial_{j}z.
\end{equation}
%The corresponding pseudomagnetic field can be evaluated from $\bl{B}\propto\nabla\times(\epsilon_{xx}-\epsilon_{yy},\,-2\epsilon_{xy})$.
We also consider a periodic scalar potential $V(\bl{r})=V_0\sum_{i=1}^{3}\cos(\bl{b}_i\cdot\bl{r}+\phi)$, which characterizes the electrostatic potential variation on the
corrugated surface in the presence of an out-of-plane displacement field.

This model serves as an example that contrasts with both Model I and I': (i) Here the pseudomagnetic field $\bl{B}(\bl{r})$ is derived purely from the out-of-plane atomic displacements in the corrugated surface, and (ii) the profiles of $\bl{B}(\bl{r})$ and $V(\bl{r})$ differ significantly (see e.g., Fig.~\ref{Fig:TareqModel_h_and_B}). Nevertheless, as will be shown explicitly in the following, this model produces qualitatively and quantitatively consistent results with the other models. 
We refer to this configuration as model II:
\begin{equation}
	\left.
	\begin{aligned}
		\epsilon_{ij}&=\frac{1}{2}\partial_{i}z\,\partial_{j}z\\
		\bl{B}&=-\frac{\sqrt{3}\hbar\beta}{2ae}\nabla\times(\epsilon_{xx}-\epsilon_{yy},\,-2\epsilon_{xy})\\
		V(\bl{r})&=V_0\sum_{i=1}^{3}\cos(\bl{b}_i\cdot\bl{r}+\phi)
	\end{aligned}
	\right\}
	\text{ Model II}
\end{equation}

\subsection{Pseudomagnetic field from the corrugated surface of model II and bulk electronic properties}

\begin{figure}[h!]
	\centering
	\includegraphics[width=4.5in]{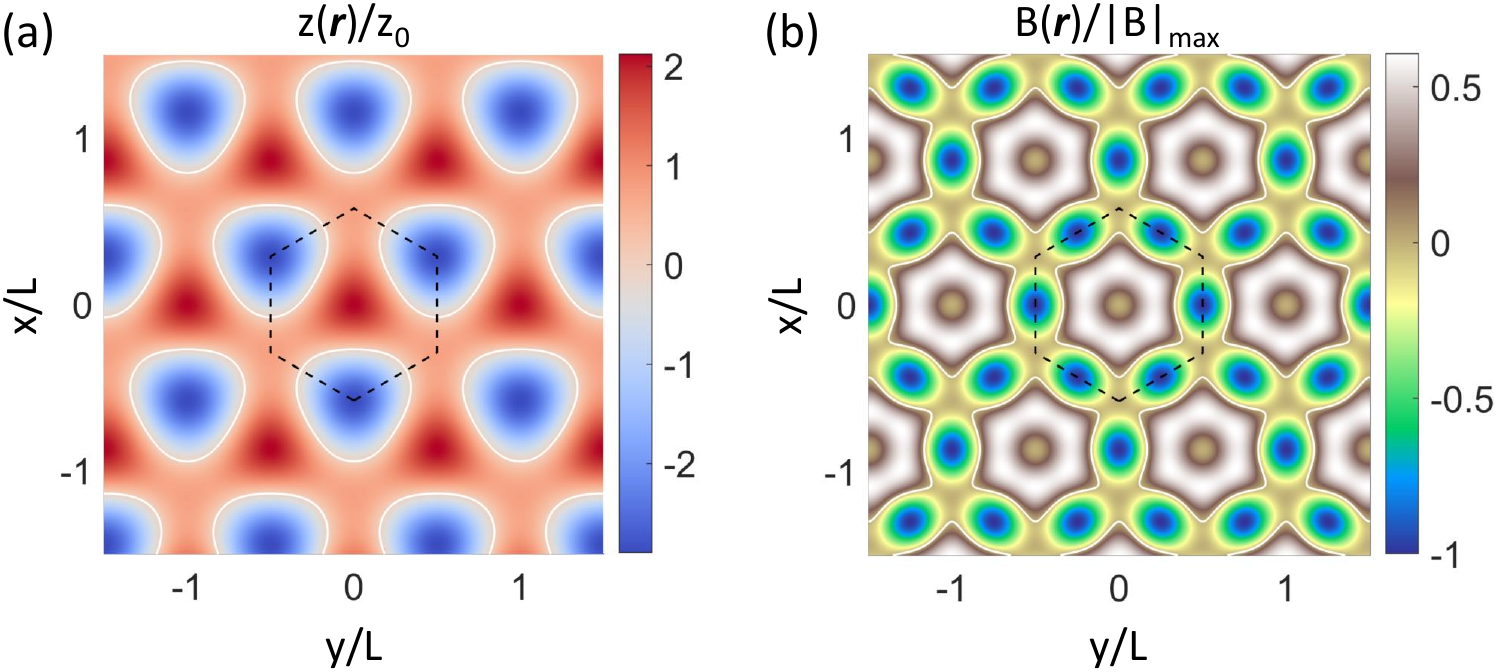}
	\caption{(a) Schematics of the strain superlattice characterized by out-of-plane corrugation $z(\bl{r})=z_0\sum_{i=1}^{3}\cos(\bl{b}_i\cdot\bl{r}+\pi/4)$. (b) The corresponding pseudomagnetic field in the $K$ valley. The white curves mark contours of $z(\bl{r})=0$ or $B(\bl{r})=0$, the black dashed lines enclose an hexagonal unit cell.}~\label{Fig:TareqModel_h_and_B}
\end{figure}

In this model, $\phi\ne n\pi/3$ with $n$ an integer is needed to obtain isolated bulk energy bands~\cite{TareqNanoLett2023}. For illustrative purposes, here we consider $\phi=\pi/4$. 
Fig.~\ref{Fig:TareqModel_h_and_B}(a) and Fig.~\ref{Fig:TareqModel_h_and_B}(b) show the distribution of $z(\bl{r})$ and $B(\bl{r})$, respectively. In contrast to model I in the main text, in this case the maxima and minima of $B(\bl{r})$ do not coincide exactly with the maxima and minima of $z(\bl{r})$.
The strain intensity $\propto z_0^2/L^2$ and the maximal value is about $100z_0^2/L^2$ (from $\epsilon_{yy}$). The pseudomagnetic field intensity $\propto z_0^2/L^3$, and the maximal field intensity $|B|_{\rm max}\approx53$~T with $z_0=0.1$~nm and $L=10$~nm.

\begin{figure}[h!]
	\centering
	\includegraphics[width=6.8in]{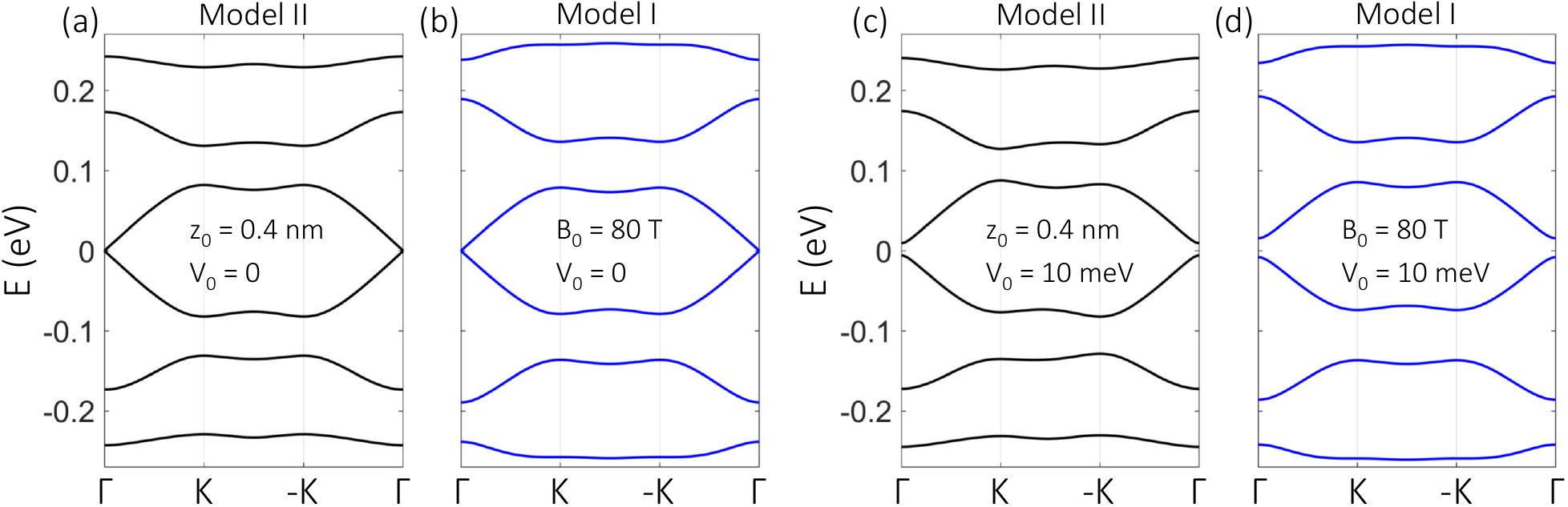}
	\caption{(a) $K$ valley energy bands of model II with $z_0=0.4$~nm and $V_0=0$. (b) $K$ valley energy bands of model II with $B_0=80$~T and $V_0=0$. (c, d) Similar results as in (a, b) but with $V_0=10$~meV. Superlattice period $L=14$~nm in all the panels.}~\label{Fig:TareqModel_ContinuumBands}
\end{figure}

Figs.~\ref{Fig:TareqModel_ContinuumBands}(a) and (c) present representative energy bands for the $K$ valley of model II, evaluated using the continuum model, in the absence or presence of the scalar potential. In comparison, Figs.~\ref{Fig:TareqModel_ContinuumBands}(b) and (d) show the bands of model I (i.e., the one in the main text), where the pseudomagnetic field and scalar potential are comparable to those for model II (maximal field intensity $|B|_{\rm max}\approx310$~T in Figs.~\ref{Fig:TareqModel_ContinuumBands}(a) and (c)). The energy bands of the two models are both qualitatively and quantitatively similar, including the band widths and gap sizes. The zero-energy gap in Fig.~\ref{Fig:TareqModel_ContinuumBands}(c) is opened by an effective staggered potential $\Delta_{\rm eff}<0$ and the first valence band has Chern number $-1$, both of these quantities are consistent with their counterparts in Fig.~\ref{Fig:TareqModel_ContinuumBands}(d).

\subsection{Scaling behavior of the bulk bands}

Similar to the discussions in Sec.~\ref{Sec.ScalingTriangularB}, now we show that the bulk energy bands exhibit some scaling behavior that is determined by the factor $\chi\propto z_0^2/L$. We illustrate this by using the continuum model in the $K$ valley: $\mathcal{H}_+=v\boldsymbol{\sigma}_+\cdot\left(\bl{p}+e\bl{A}\right)$, where only the vector potential is considered at the moment. The pristine part can be rewritten as
\begin{equation}
	H_0=v\boldsymbol{\sigma}_+\cdot\bl{p}=E_0\boldsymbol{\sigma}_+\cdot\tilde{\bl{k}},
\end{equation}
where $E_0=\hbar v|\bl{b}_i|=\hbar v\frac{4\pi}{\sqrt{3}L}$ characterizes the strength of the pristine kinetic energy and $\tilde{\bl{k}}=\bl{k}/|\bl{b}_i|$ is dimensionless. The part contributed by the vector potential can be rewritten as
\begin{equation}
	H_v=ev \boldsymbol{\sigma}_+\cdot\bl{A}=-g_v\boldsymbol{\sigma}_+\cdot(\epsilon_{xx}-\epsilon_{yy},\,-2\epsilon_{xy})=-g_v\frac{z_0^2}{L^2}\boldsymbol{\sigma}\cdot(\tilde{\epsilon}_{xx}-\tilde{\epsilon}_{yy},\,-2\tilde{\epsilon}_{xy}),
\end{equation}
where $g_v=\frac{\sqrt{3}\hbar v\beta}{2a}\approx6.953$ eV, and $\tilde{\epsilon}=\epsilon L^2/z_0^2$. In these forms, the Hamiltonian becomes
\begin{equation}
	\mathcal{H}_+=E_0\boldsymbol{\sigma}_+\cdot\left(\tilde{\bl{k}}-\frac{g_v}{E_0}\frac{z_0^2}{L^2}\tilde{\bl{A}}\right)=E_0\boldsymbol{\sigma}_+\cdot\left(\tilde{\bl{k}}-\chi\tilde{\bl{A}}\right),
\end{equation}
where $\chi=\frac{g_v}{E_0}\frac{z_0^2}{L^2}\propto z_0^2/L$ is the factor that characterizes the ratio of the vector potential energy to the kinetic energy, and we have also defined $\tilde{\bl{A}}=(\tilde{\epsilon}_{xx}-\tilde{\epsilon}_{yy},\,-2\tilde{\epsilon}_{xy})$. The eigenenergy problem of solving $\mathcal{H}_+\psi=E\psi$ is equivalent to
\begin{equation}
	\boldsymbol{\sigma}_+\cdot\left(\tilde{\bl{k}}-\chi\tilde{\bl{A}}\right)\psi=\tilde{E}\psi,
\end{equation}
where $\tilde{E}=E/E_0$.
From this, it becomes clear that if we measure the wave vector in units of $|\bl{b}_i|$, and energy in units of $E_0$, the factor $\chi\propto z_0^2/L$ is the only independent tuning parameter in the model. Generally, a larger $\chi$ leads to narrower bands.

In the presence of a small periodic scalar potential and/or constant staggered potential, as discussed in the main text, a gap opens at the Dirac point, while the dispersions at higher energies are modestly affected [also compare Fig.~\ref{Fig:TareqModel_ContinuumBands}(a) with Fig.~\ref{Fig:TareqModel_ContinuumBands}(c)].

In later tight-binding calculations (Fig.~\ref{Fig:S_model2}), we consider $z_0=0.25$~nm and a small period $L=21a=5.165$~nm to reduce computational cost, which lead to $\chi=0.0121$~nm$^{-1}$. In Figs.~\ref{Fig:TareqModel_ContinuumBands}(a, c), experimentally realistic values of $z_0=0.4$~nm and $L=14$~nm yield $\chi\approx0.01143$~nm$^{-1}$, close to that in the tight-binding approach. 
%After rescaling the energy by $E_0\propto 1/L$, one finds consistent energy bands from tight-binding and continuum models.
Results from the tight-binding and continuum models are consistent (up to the scaling).

\subsection{Edge states in a nanoribbon}

In the presence of strain, the hopping energy is modified as $t_{ij}=t_0e^{-\beta\left( \frac{d_{ij}}{a_{0}}-1 \right)}$, where $d_{ij}$ is the distance between two nearest-neighbor sites located at $(\bl{r}_i,z(\bl{r}_i))$ and $(\bl{r}_j,z(\bl{r}_j))$. Here $\bl{r}_i$ is understood to be in the $xy$ plane defined by the lattice sites of pristine graphene.

\begin{figure}[h!]
	\centering
	\includegraphics[width=4.5in]{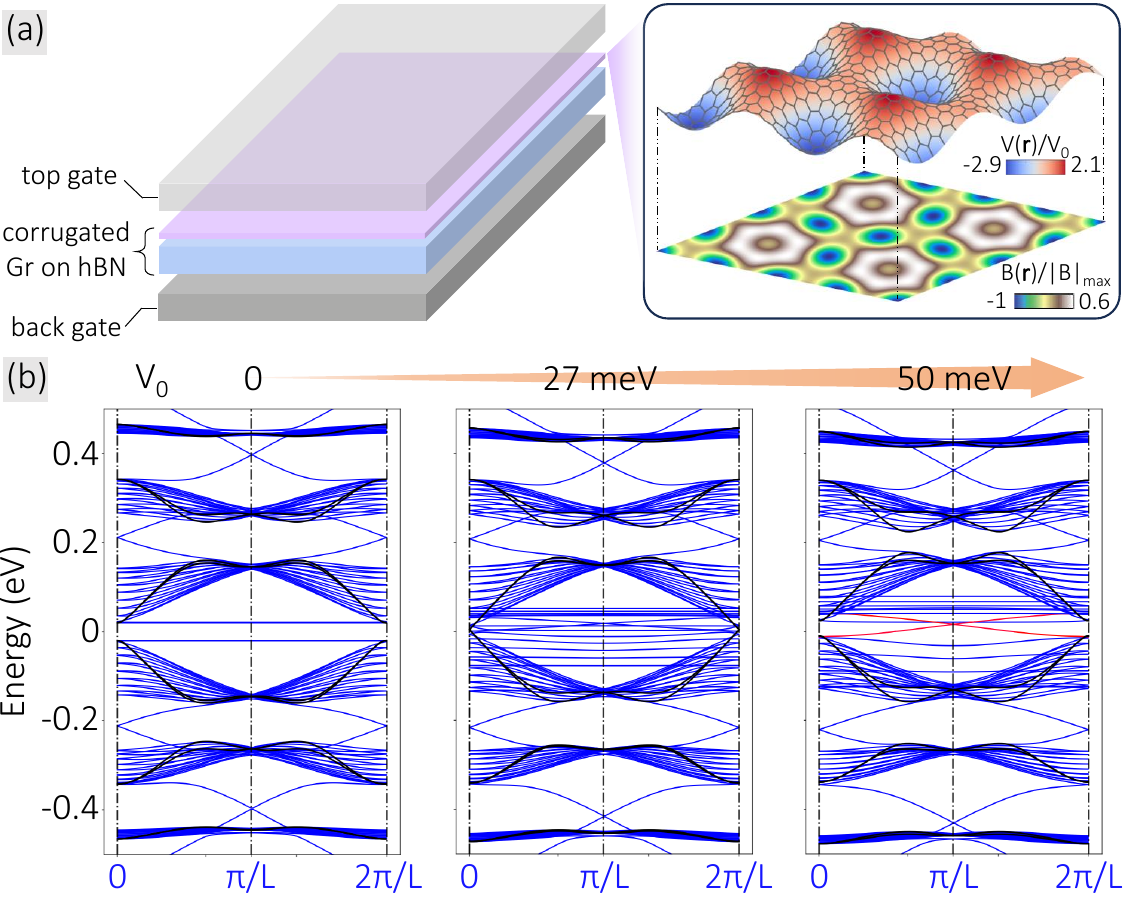}
	\caption{\textbf{Model II}. (a) Schematics of periodically corrugated graphene on hBN with height profile $z(\bl{r})=z_0\sum_{i=1}^{3}\cos(\bl{b}_i\cdot\bl{r}+\frac{\pi}{4})$. An out-of-plane displacement field can be applied to generate the periodic scalar potential $V(\bl{r})=V_0\sum_{i=1}^{3}\cos(\bl{b}_i\cdot\bl{r}+\frac{\pi}{4})$. The pseudomagnetic field $B(\bl{r})$ is obtained from the strain tensor contributed by $z(\bl{r})$ as described in the text. The hBN introduces a staggered potential $\Delta\sigma_z$ onto graphene. (b) Evolution of the energy bands of a zigzag nanoribbon of width $6.1L=31.51$~nm (blue curves) with respect to $V_0$ by tuning the displacement field. The black curves represent the bulk bands along the same high symmetry path as in Figs.~1(b, c) of the main text. $z_0=0.25$~nm and $\Delta=20$~meV have been used.}~\label{Fig:S_model2}
\end{figure}

Fig.~\ref{Fig:S_model2}(b) presents the energy bands of Model II with $h_0=0.25$~nm, $\Delta=20$~meV and three different $V_0$ values. In this model, $V_0>0$ also leads to effective staggered potential $\Delta_{\rm eff}<0$. As the profiles of $B(\bl{r})$ and $V(\bl{r})$ are even more different here, the resultant $\Delta_{\rm eff}$ has a smaller magnitude compared to those in both Model I and I', thus a larger $V_0$ is needed in Fig.~\ref{Fig:S_model2}(b) to close the zero-energy gap initially introduced by $\Delta$. Importantly, for a large $V_0$, propagating edge states emerge with energies in the zero-energy gap [Fig.~\ref{Fig:S_model2}(b) third panel].

\end{document}